\documentclass[preprint,12pt,a4paper]{article}
\usepackage{setspace}
\usepackage{appendix}
\usepackage{graphicx}  % needed for figures
\usepackage{dcolumn}   % needed for some tables
\usepackage{bm}        % for math
\usepackage{amssymb}   % for math
\usepackage{amsmath}
\usepackage{array}
\usepackage{epstopdf}
\usepackage{mathtools}
\usepackage[T1]{fontenc}  % access \textquotedbl
\usepackage{textcomp}     % access \textquotesingle
\usepackage{mathtools,slashed}
\usepackage{graphics}
\usepackage{epsfig}
\usepackage{psfrag}
\usepackage{xcolor,colortbl}
\usepackage[utf8]{inputenc}
\usepackage{scrextend}
\DeclareMathAlphabet{\pazocal}{OMS}{zplm}{m}{n}
\usepackage{amsfonts}
\usepackage{amsthm}
\usepackage{float}
\usepackage{textcomp}
\usepackage[section]{placeins}
\usepackage[caption=false]{subfig}
\usepackage{subfig}
\usepackage{tabularx}
\usepackage{hyperref}
\usepackage{ulem}
\usepackage{multirow}
\newcolumntype{g}{>{\columncolor{Gray}}c}
\newcolumntype{w}{>{\columncolor{Gray2}}c}
\definecolor{Gray}{gray}{0.95}
\definecolor{Gray2}{gray}{0.98}

\def\be{\begin{equation}}
\def\ee{\end{equation}}
\def\bea{\begin{eqnarray}}
\def\eea{\end{eqnarray}}

\begin{document}
\begin{center}
\baselineskip 20pt 
%{\Large\bf Smooth Hybrid Inflation  
%	\\ \vspace{0.2cm} in $SU(5) \times U(1)_{\chi}$ Super-GUT}
{\Large\bf Primordial Black Holes and Gravitational Waves  in Hybrid Inflation with Chaotic Potentials}
\vspace{1cm}

{\large 
	Waqas Ahmed$^{a}$ \footnote{E-mail: \texttt{\href{mailto:waqasmit@hbpu.edu.cn}{waqasmit@hbpu.edu.cn}}},
	 M. Junaid$^{b}$ \footnote{E-mail: \texttt{\href{mailto:umer@udel.edu}{mjunaid@ualberta.ca}}} 
	 and Umer Zubair$^{c,d}$ \footnote{E-mail: \texttt{\href{mailto:umer@udel.edu}{umer@udel.edu}}; \texttt{\href{mailto:uzubair@sju.edu}{uzubair@sju.edu}}}
} 
\vspace{.5cm}

{\baselineskip 20pt \it
	$^{a}$ \it
	School of Mathematics and Physics, Hubei Polytechnic University, \\
	 Huangshi 435003,
	China \\
	\vspace*{6pt}
	$^{b}$National Centre for Physics, Islamabad, Pakistan \\
	\vspace*{6pt}
	$^c$Department of Physics and Astronomy,  \\
	University of Delaware, Newark, DE 19716, USA \\
	\vspace*{6pt}
	$^d$Department of Physics, \\Saint Joseph’s University, Philadelphia, PA 19131, USA 
	\vspace{2mm} }

\vspace{1cm}
\end{center}

\begin{abstract}
We study the formation of primordial black hole (PBH) dark matter and the generation of scalar induced secondary gravitational waves (SIGWs) in a non-supersymmetric model of hybrid inflation with chaotic (polynomial-like) potential, including one-loop radiative corrections. A radiatively corrected version of these models is entirely consistent with Planck's data. By adding non-canonical kinetic energy term in the lagrangian, the inflaton experiences a period of ultra-slow-roll, and the amplitude of primordial power spectrum is enhanced to $O(10^{-2})$. The enhanced power spectra of primordial curvature perturbations can have both sharp and broad peaks. %In the enhancement mechanism we explore two possible extensions by employing a Guassian and a step size kinetic function. 
A wide mass range of PBHs is realized in our model with the frequencies of scalar induced gravitational waves ranged from nHz to kHz. We present several benchmark points where the PBH mass generated during inflation is around $(1 - 100) \, M_{\odot}$, $(10^{-9} - 10^{-7}) \, M_{\odot}$ and $(10^{-16} - 10^{-11}) \, M_{\odot}$. The PBHs can make up most of the dark matter with masses around  $(10^{-16} - 10^{-11}) \, M_{\odot}$ and $(1 - 100) \, M_{\odot}$, and their associated SIGWs can be probed by the upcoming ground and space-based gravitational wave (GW) observatories. The evidence of stochastic process recently reported by NANOGrav may be interpreted as SIGWs associated with the formation of PBHs. These SIGWs may also be tested by future interferometer-type GW observations of SKA, DECIGO, LISA, BBO, TaiJi, TianQin, CE and ET.
\end{abstract}

\section{Introduction}
From astronomical and cosmological observations, there is convincing evidence that $85 \%$ of the matter in the Universe is in the form of cold, non-baryonic dark matter (DM)\cite{Bertone:2016nfn}. The study of Primordial Black Holes (PBHs) dates back to the 1960's and 70's \cite{Zeldovich:1966, Hawking:1971ei, Carr:1974nx}, and shows that the PBHs may form due the collapse of large over-densities in the early universe. The early universe may contain regions with high densities at small scales that can trigger gravitational collapse to form PBHs. This PBH production can be tested through their effects on a variety of cosmological and astronomical processes, and therefore, can serve as an inspiring tool to probe physics in the very early Universe \cite{Khlopov:2008qy, Sasaki:2018dmp}. In particular, PBH could be a potential candidate for (a fraction of) dark matter (DM), which has drawn a lot of attention \cite{Carr:2016drx, Carr:2018poi}.

PBHs are non-baryonic, as they form before matter-radiation equality. The PBHs with masses $\lesssim 10^{15}$ g would have evaporated by now emitting Hawking radiation \cite{Hawking:1974rv}. The emitted particles may impact the gamma-ray background \cite{MacGibbon:1991vc} and the abundance of light elements produced by the big bang nucleosynthesis (BBN) \cite{Carr:2009jm}. The PBHs with masses greater than $10^{15}$ g, on the other hand may survive up to the present epoch and are expected to be constrained by their gravitational effects, such as gravitational lensing \cite{Niikura:2019kqi}, dynamical effects on baryonic matter \cite{Carr:1997cn}, or the fast radio bursts created by mergers of charged PBHs \cite{Deng:2018wmy}. 

The Laser Interferometer Gravitational-Wave Observatory (LIGO), the Scientific Collaboration and the Virgo Collaboration have detected several events of GWs coming from the merger of black holes (BHs) \cite{GW1,GW2,GW3,GW4,GW5}. Recently, the North American Nano hertz Observatory for Gravitational Wave (NANOGrav) Collaboration \cite{Arzoumanian:2020vkk} has published an analysis of the 12.5 yrs pulsar timing array (PTA) data, where strong evidence of a stochastic process with a common amplitude and a common spectral slope across pulsars was found. These two observations moved the physicists attention toward the gravitational waves (GWs) generated by PBH-PBH mergers \cite{Sasaki:2016jop, Mandic:2016lcn, Wang:2016ana}, as well as the scalar-induced GWs from the enhanced primordial density perturbations associated with PBH formation \cite{Baumann:2007zm, Ananda:2006af, Kohri:2018awv, Bartolo:2018rku, Cai:2019jah, Cai:2018dig}. The GWs survey shall be a promising window to reveal the physical processes of PBH formations.

The scalar induced gravitational waves (SIGWs) associated with the formation of PBHs may be the source of  NANOGrav signal \cite{DeLuca:2020agl}, or the GWs detected by LIGO/Virgo. In order to produce PBHs in the radiation era from the gravitational collapse of overdense regions, it is required that the density of overdense regions exceed the threshold value at the horizon re-entry. The initial conditions for these overdense regions are produced during the inflationary era. To produce the desired abundance of PBHs one needs the primordial scalar power spectrum at small scales to be enhanced to $P_\zeta (k > 1) \sim {O(0.01)}$. This condition is also required to explain the NANOGrav signal if it is regarded as a SIGW. On the other hand, the constraint on the amplitude of power spectrum at large scales from the cosmic microwave background (CMB) anisotropy measurements from Planck \cite{Planck:2018jri} is $P_\zeta(0.05)\sim {O(10^{-9})}$. To produce enough abundance of PBH dark matter (DM) and SIGWs measurable by NANOGrav, the amplitude of the power spectrum at small scales should be enhanced at least seven orders of magnitude to reach the threshold value.

In this paper, we study the non-supersymmetric model of hybrid inflation and the formation of primordial black hole (PBH) dark matter with the generation of their associated scalar induced secondary gravitational waves (SIGWs). In order to produce the required abundance of PBHs, we enhance the power spectrum using the mechanism discussed in \cite{Lin:2020goi} and extend the mechanism to study two more possibilities. This mechanism relies on the incorporation of a non-canonical kinetic term with a function $G(\phi)$ which exhibits a peak at some value of the field $\phi_p$. Beyond this point, $G(\phi)$ falls exponentially, suppressing the scalar perturbations to the value observed today. With the extended mechanism, the predictions of our model are in much better agreement with the experimental bounds. Several other enhancement mechanisms are also discussed in the literature as well. The enhancement by ultra-slow-roll inflation with an inflection point is discussed in \cite{Lu:2019sti}. The other possibility is fine-tuning the model parameters while keeping the total number of $e$-folds around 50-60  \cite{Garcia-Bellido:2017aan}. 

After the formation of PBHs, the enhanced power spectrum, at small scales, induces secondary GWs after the horizon re-entry during the radiation-dominated epoch \cite{Matarrese:1997ay}. These SIGWs have a vast range of frequencies and consist of a stochastic background that can be detected by ground and space-based future GW detectors such as Square Kilometre Array (SKA) \cite{Smits:2008cf}, North American Nanohertz Observatory for Gravitational Waves (NANOGrav) \cite{Arzoumanian:2020vkk}, Einstein Telescope (ET) \cite{Punturo:2010zz}, Cosmic Explorer (CE)\cite{LIGOScientific:2016wof}, Laser Interferometer Gravitational-Wave Observatory (LIGO) O5 \cite{LIGOScientific:2019vic} , Laser Interferometer Space Antenna (LISA) \cite{Ferdman:2010xq,LISA:2017pwj}, Deci-hertz Interferometer Gravitational wave Observatory (DECIGO)\cite{Seto:2001qf}, Big Bang Observer (BBO) \cite{Corbin:2005ny}, TaiJi \cite{Hu:2017mde}, TianQin \cite{TianQin:2015yph} and Atomic Experiment for Dark Matter and Gravity Exploration in Space (AEDGE)\cite{AEDGE:2019nxb}.

The layout of the paper is as follows. In Sec. \ref{sec_1} we present the non-supersymmetric hybrid inflation model. The inflation setup and PBH production is described in Sec. \ref{sec_2}. PBH abundance is computed in Sec. \ref{sec_3} while the generation of SIGWs is studied in Sec. \ref{sec_4}. Our conclusions are summarized in Sec. \ref{sec_5}.

\section{Description of the Model} \label{sec_1}

The scalar potential of hybrid inflation (HI) can be expressed as a combination of Higgs potential $V(\chi)$ and inflaton potential $\delta V(\phi)$ with an additional term, $g^{2}\chi^{2}\phi^{2}$, which represents the interaction between the Higgs field $\chi $ and inflaton  $\phi$. The tree-level hybrid inflation potential, therefore, can be written as 
\begin{equation} \label{GP}
V(\chi,\phi)=\kappa^{2}\left(M^{2}-\frac{\chi^{2}}{4}\right)^{2}+\frac{g^{2}\chi^{2}\phi^{2}}{4}+\delta V(\phi),
\end{equation}
where $\delta V(\phi)$ is the inflaton potential and is taken to be a chaotic polynomial-like potential, i.e., $\delta V(\phi) = \lambda_{p}\phi^{p}$ with  $ p>0 $. Here, the role of the interaction term is to generate an effective (squared) mass, 
\begin{equation}
m^{2}_{\chi} = -\kappa^{2}M^{2}+\frac{g^{2}\phi^{2}}{2}
= \frac{g^{2}}{2}\left( \phi^{2} - \phi_c^{2}\right), \text{ with } \phi_{c} \equiv \frac{\sqrt{2}\kappa M}{g},
\end{equation}
for the $\chi$ field in the $ \chi=0 $ direction. This direction is a local minimum for $\phi > \phi_{c} =\frac{\sqrt{2}\kappa M}{g}$ and can be used for inflation 
with effective single field potential given by
\begin{equation}\label{EP}
V(\phi)= \kappa^{2}M^{4} + \delta V(\phi) = V_{0}+\lambda_{p}\phi^{p},
\end{equation}
where $V_{0}=\kappa^{2}M^{4} $. The chaotic potential, here, provides the necessary slope for the slow-roll inflation in the otherwise flat-valley. We consider suitable initial conditions for inflation to occur only in the $\chi=0 $ valley until $\phi = \phi_{c}$ is reached where inflation is terminated abruptly, followed by a waterfall phase transition. 

We now tend to include one-loop radiative corrections, as the tree level predictions are not consistent with the Planck 2018 results \cite{Ahmed:2014cma}. The radiative corrections arise from the possible coupling of inflaton with other fields. These couplings can contribute to the reheating process in order to recover the hot big bang initial conditions. The corrections arising from the coupling of inflaton to fermions or bosons may be termed as fermionic or bosonic radiative corrections. The one-loop radiative corrections to $V(\phi)$ in the inflationary valley can be found from the following form of Coleman-Weinberg formula \cite{Coleman:1973jx}, 
\be \label{loop}
V_{\text{1-loop}} =  A \, \phi^{4}\ln \left(\frac{ \phi}{\phi_{c}}\right), 
\ee
where $A<0$ ($A>0$) for fermonic (bosonic) radiative corrections. The fermionic radiative corrections have already been seen to play an important role for the chaotic inflation driven by the quadratic and the quartic potentials \cite{NeferSenoguz:2008nn}. The fermionic radiative corrections generally reduce both $r$ and $n_s$ in the chaotic inflation. In the following, we study the effect of fermionic radiative corrections on the tree-level predictions and  compare them with the Planck's latest bounds on $r$ and $n_s$. 

Using Eqs.\eqref{EP} and \eqref{loop}, the one-loop radiatively corrected hybrid inflationary (RCHI) potential can be written as,
\begin{equation}
V = V_{0} + \lambda_{p} \phi^{p} -  A \, \phi^{4} \ln \left(\frac{\phi}{\phi_{c}}\right). \label{Vphi}
\end{equation}

In order to discuss the predictions of the model, some discussion of the effective number of independent parameters is in order. Apart from the parameter $\lambda_p$ of the chaotic potential, the fundamental parameters of the potential in Eq.~(\ref{GP}) are $\kappa$, $g$ and $M$, which can be reduced to $V_{0}$ and $\phi_{c}$ for the effective potential in Eq.~(\ref{EP}). We, however, take $V_0$ and $\kappa_c \equiv g^{2}/\kappa$ as the effective independent parameters with $\phi_{c} = \sqrt{2V_{0}^{1/2}/\kappa_c}$. With this choice we can develop a simple correspondence for the supersymmetric hybrid inflation for which $\kappa_c = g = \kappa$ \cite{Dvali:1994ms}. 

\begin{table}[!htb]
	\setlength\extrarowheight{5pt}
\centering
\resizebox{\textwidth}{!}{
\begin{tabular}{g|c|c|c|c|c|c|c } 
 \hline \hline \rowcolor{Gray}
 Model & $V_0$ & $\lambda_p$ & $A$ & $\phi_c$ & $\phi_*$ & $n_s$ & $r$ \\ 
 \hline 
  $p = 1$    & $5.45\times10^{-11}$ & $9.76\times10^{-13}$ & $8.79\times10^{-14} $& 0.1215 & 0.63 & 0.965 & 0.00176\\
 $p = 2$    &  $5.45\times10^{-11}$ & $9.33\times10^{-13}$ & $1.67\times10^{-13}$ & 0.1215 & 0.76 & 0.965 & 0.00173\\
 $p = 2/3$ & $5.45\times10^{-11} $& $1.31\times10^{-12}$ & $2.96\times10^{-14}$ & 0.1215 & 0.788 & 0.965& 0.0018\\
\hline \hline
\end{tabular}}
\caption{Hybrid inflation with chaotic potential parameters $V_0$, $\lambda_p$, $A$ and $\phi_c$ for  $p = 1$, $p = 2$ and $p = 2/3$. $\phi_*$ corresponds to the value of $\phi$ at the pivot scale $k_* = 0.05 \, \text{Mpc}^{-1}$. The scalar spectral index $n_s$ and power spectra $P_\zeta(k_*)=2.15\times 10^{-9}$ are evaluated at the pivot scale $k_*$ for the three chaotic potentials. }
\label{tb1}
\end{table}

\section{Inflation and PBH Production} \label{sec_2}

PBHs are formed from the gravitational collapse of over-dense regions when their density contrasts at the horizon re-entry during radiation domination exceeding the threshold value. The overdense regions may be seeded from the primordial curvature perturbations generated during inflation. The feasible way to produce enough abundance of primordial black hole (PBH) dark matter (DM) is by enhancing the amplitude of the power spectrum at least seven orders of magnitude to reach the threshold at small scales. We employ the enhancement mechanism of the power spectrum at small scales proposed in Refs \cite{Lin:2020goi} using kinetic or K/G inflation. The kinetic inflation is defined when inflaton field's kinetic part is coupled to inflaton field function $K(\phi)=1+G(\phi)$ as \cite{Kobayashi:2010cm}
\begin{eqnarray}
S = \int d^4x \sqrt{-g} \left ( m_p^2\frac{R}{2} - \frac{1}{2} g^{\mu\nu} K(\phi) \partial_{\mu}\phi \partial_{\nu}\phi +V(\phi) \right),\notag
\end{eqnarray}
where %$X=-g_{\mu\nu}\nabla^{\mu}\phi\nabla^{\nu}\phi/2$, 
$m_{p}=1/\sqrt{8\pi G}=1$. The background equations of motion are
\begin{eqnarray}
\label{Eq:eom1}
H^2 &=& \frac{1}{3} \left(K(\phi)\frac{\dot{\phi}^2}{2} + V(\phi) \right),\\
\dot{H}&=&-K(\phi)\frac{\dot{\phi}^2}{2} ,\\
\ddot{\phi}&=& -3H\dot{\phi} - \frac{V_{,\phi}+K_{,\phi}\dot{\phi}^2 /2}{K(\phi)},
\end{eqnarray}
where $K_{,\phi}=dK(\phi)/d\phi$. These equations can be written in terms of derivatives with respect to number of efolds $n=\ln(a)$ as follows
\begin{equation}
	H^2 = \frac{2V}{6-K(\phi)\phi^2_{,n}},
\end{equation}
\begin{eqnarray}
%H^2&=&\frac{2V}{6-K(\phi)\phi^2_{,n}},\\
H_{,n} &=& -K(\phi) H\frac{\phi_n^2}{2} ,\\
\phi_{,nn} &=& -\left(3+\frac{H_{,n}}{H}\right)\left(\phi_{,n}+\frac{V_{,\phi}}{K(\phi)V}\right)-\frac{\phi_{,n}^2K_{,\phi}}{2K(\phi) }.
\end{eqnarray}
The slow-roll parameters are defined as
\begin{eqnarray}
\epsilon_1&=&-\frac{\dot{H}}{H^2}=-\frac{H_{,n}}{H},\\ 
\epsilon_2&=&-\frac{\ddot{\phi}}{H\dot{\phi}}=-\left( \frac{\phi_{,nn}}{\phi_{,n}}+\frac{H_{,n}}{H}\right),\\
\epsilon_K&=&\frac{\dot{\phi}K_{,\phi}}{H K(\phi)}=\frac{\phi_{,n}K_{,\phi}}{K(\phi)},
\end{eqnarray}
where the last slow roll parameter is specific for Kinetic Inflation. The slow-roll inflation is realized when $|\epsilon_i|\ll1$, where $i=1,2,K$.

The second order action $S^{(2)}$ in perturbation theory of the curvature perturbation $\zeta$, for Kinetic Inflation is \cite{Garriga:1999vw},
\begin{equation}
S^{(2)}=\frac{1}{2}\int d\tau d^3x\tilde{z}^2 K(\phi)[(\zeta')^2-(\vec{\nabla}\zeta)^2],\label{eq:cp}
\end{equation}
where $\tilde{z}=a(t)\dot{\phi}/H$ and $\zeta'$ represents derivative with respect to the conformal time $\tau=\int dt/a(t)$.
%Since the sound speed for the scalar mode is $c_s^2=1$,
%so there is no problem with ghost and gradient instabilities.
Using $z=\sqrt{K}\tilde{z}$ in the quadratic action \eqref{eq:cp} and varying with respect to curvature perturbation $\zeta_k=v_k/z$ and $v_k$ in the Fourier space, we obtain the famous Mukhanov equations for scalar perturbations
\begin{eqnarray}
v''_k&+&\Bigl(k^2-\frac{z''}{z}\Bigl)v_k=0,\\
\zeta''_k&+&2\frac{z'}{z}\zeta'_k+k^2 \zeta_k=0 ~.
\label{eq:1}
\end{eqnarray}
The above mode equation for $\zeta_k$ can be written in terms of derivatives with respect to number of e-folds $n$ as follows
\begin{eqnarray}
\zeta_{k, nn} &+& \left(3 + \frac{H_{,n}}{H}+\frac{2\phi_{,nn}}{\phi_{,n}} +\frac{\phi_{,n}K_{,\phi}}{K(\phi)} \right)\zeta_{k, n} +\left(\frac{k}{aH}\right)^2 \zeta_{k}= 0,\\
\zeta_{k, nn} &+& \left(3 + \epsilon_1-2\epsilon_2+\epsilon_K \right)\zeta_{k, n} +\left(\frac{k}{aH}\right)^2 \zeta_{k}= 0~,\label{eq:2}
\end{eqnarray}
where we have used
\begin{equation*}
	z=\sqrt{K(\phi)}a\phi_{,n}\, ,
\end{equation*} 
and 
\begin{equation*}
	\frac{z'}{z}=a H\left(1+\frac{\phi_{,nn}}{\phi_{,n}}+\frac{\phi_{,n}K_{,\phi}}{2K(\phi)}\right).
\end{equation*} 
Numerically it is more convenient to solve the equation for $\zeta_k$ rather than standard Mukhanov variable $v_k$, as the former yields stable results.

Solving the mode Eq. \eqref{eq:1} with Bunch Davies vacuum\cite{TasiCosmology2018}, we obtain the scalar power spectrum on super horizon scales $|k\tau|\ll 1$,
\begin{equation}
\label{eq:ps}
P_{\zeta}(k)=\frac{k^3|\zeta_k|^2}{2\pi^2}=\frac{H_*^2}{8\pi^2 \epsilon_1^*},%\simeq\frac{K(\phi_*)V^3(\phi_*)}{12\pi^2 m_p^6 V_{,\phi}^2},
\end{equation}
where $_*$ marks the horizon crossing values for each mode $k$. In the slow-roll approximation \footnote{It should be noted that in our numerical results, we have used exact form of Eq. \eqref{eq:ps} without the slow-roll approximation.}, the scalar power spectrum can be estimated  as,
\begin{equation}
	P_\zeta(k) \simeq \frac{V^3(\phi_{*})}{12 \pi^2 V_{,\phi}^2} K (\phi_*).
\end{equation} 
The scalar spectral index $n_s$, tensor to scalar ratio $r$ and the total number of e-folds $N$ for KG inflation at the pivot scale are given by
\begin{eqnarray}
\label{nseq1}
n_s&=&1+\frac{1}{K}\left(2\eta_V-6\epsilon_V-
\sqrt{2\epsilon_V}\frac{K_{,\phi}}{K}\right),\\
\label{req1}
r&=&\frac{P_T}{P_{\zeta}}\simeq 16\epsilon_V/K, \\
N&=&\int_{\phi_e}^{\phi_*} \frac{d\phi}{\sqrt{2\epsilon_1}}\simeq \int_{\phi_e}^{\phi_*} d\phi \frac{\sqrt{K} V}{V_{,\phi}},
\end{eqnarray}
where $\epsilon_V=(V_{,\phi}/V)^2/2$, $\eta_V=V_{,\phi\phi}/V$ and $N$ is the total number of e-folds.

The CMB power spectrum as reported by Planck 2018 \cite{Planck:2018jri} is $P_{\zeta}(k^*) = 2.15\times 10^{-9}$ at the pivot scale $k^* = 0.05 \, \text{Mpc}^{-1}$. For the PBH production, the power spectrum needs to be enhanced to $P_{\zeta}(k) \simeq 10^{-2}$ at small scales $k > 10^2 \, \text{Mpc}^{-1}$. This enhancement in the power spectrum can be achieved using $K(\phi)=1+G(\phi)$, as some kind of peak function.  We employ a polynomial peak function $K_q(\phi)$, from \cite{Lin:2020goi} and introduce two functions; a Gaussian peak function $K_g(\phi)$ and a step function $K_s(\phi)$,
\begin{eqnarray}
K_g(\phi) &=& 1+ h~ e^{-\frac{(\phi-\phi_p)^2}{2 w^2}},\label{kg}\\
K_q(\phi) &=& 1+ \frac{h}{\sqrt[q]{1+\frac{\left|\phi-\phi_p\right|^q}{w^q}}},\label{kq}\\
K_s(\phi) &=& 1+ \left\{
\begin{array}{cc}
h & \left(\phi_p - w/2\right) < \phi < \left(\phi_p + w/2\right) \\
0 & \phi > \left(\phi_p + w/2\right),~\phi < \left(\phi_p - w/2\right)
\end{array}\right.  \label{ks}
\end{eqnarray}
\begin{figure}[t]
	\centering 	\includegraphics[width=10.0cm]{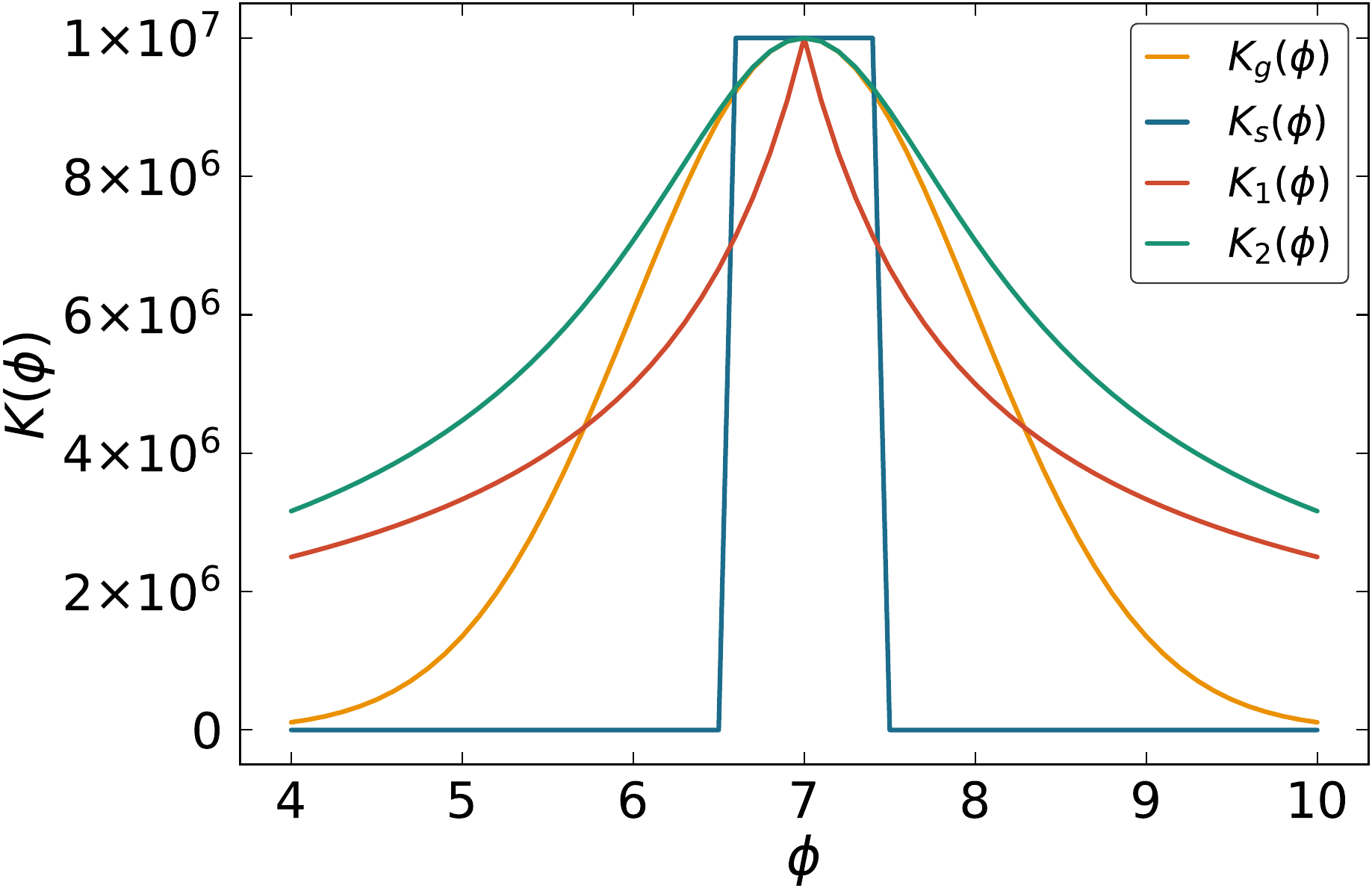}
	\caption{Gaussian peak function $K_g(\phi)$, polynomial power function $K_q(\phi)$, with $q=1,2$, and a step function $K_s(\phi)$ as a function of $\phi$ for $h=10^7$ and $w=1$.}
	\label{Kphi}
\end{figure}
\begin{figure}[!htb]
	\centering \includegraphics[width=10.0cm]{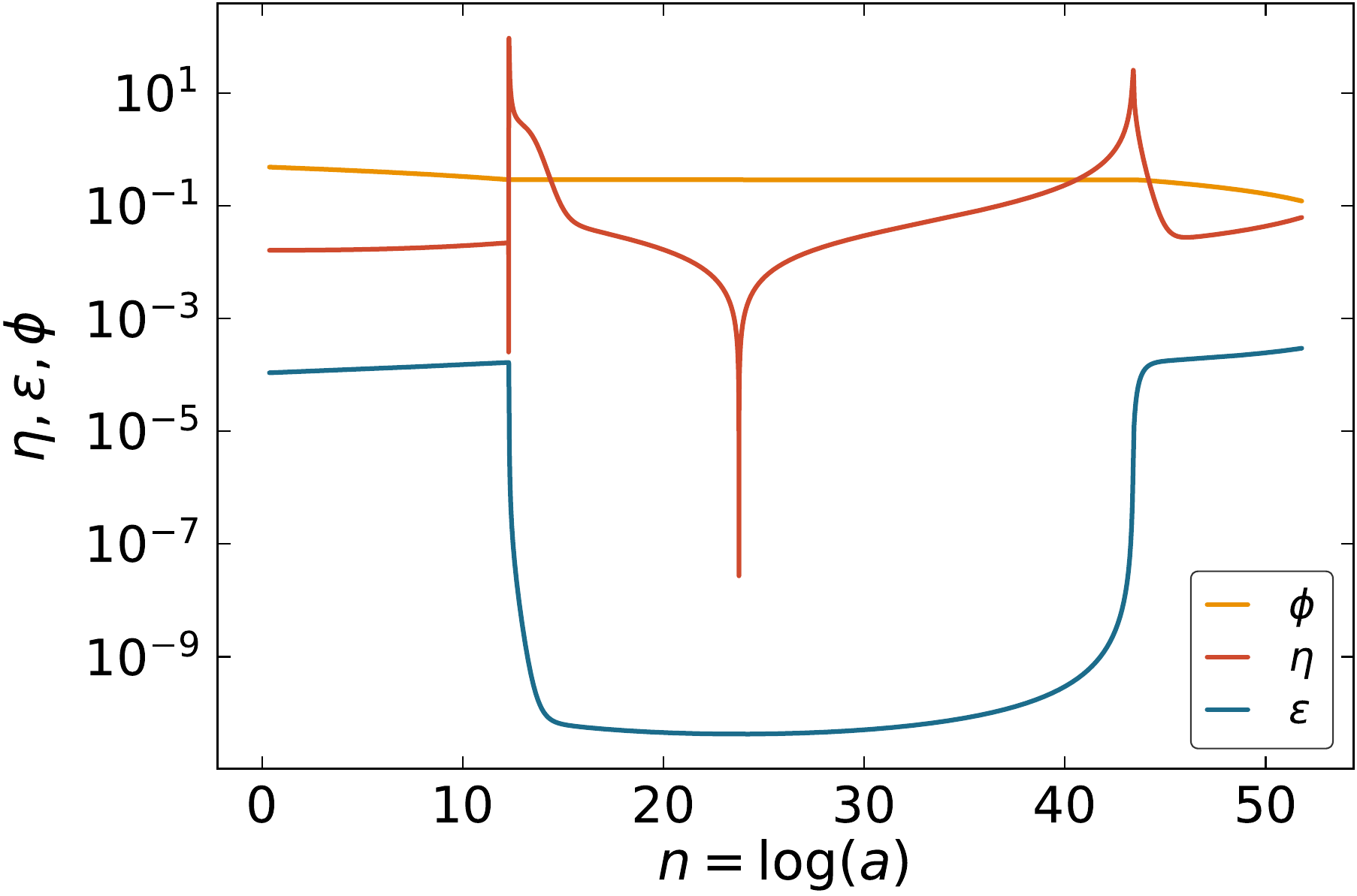}
	\caption{Behavior of background parameters $\phi$, $\epsilon$ and $|\eta|$  with respect to the number of $e$-folds $n=\log(a)$. The curves are drawn for Gaussian peak function $K_g(\phi)$. The numerical values of the relevant parameters are given in Table \ref{tb3}.}
	\label{EEP}
\end{figure}
\begin{table}[t]
	\setlength\extrarowheight{5pt}
	\centering
	\resizebox{\textwidth}{!}{
	\begin{tabular}{g|c|c|c|c|c|c} 
		\hline \hline \rowcolor{Gray}
	Peak function &	Model & $h$ & $w$ & $\phi_p$ & $Y_{\text{pbh}}^{\text{peak}}$  & $\Omega_{\text{GW}}^{\text{peak}}$  \\ 
		\hline 
		 	& $p = 1$  & $4.75\times 10^{3}$ & $7.7\times 10^{-5}$ & 0.24 & 0.032 & $2.64\times 10^{-8}$\\
	 	& $p = 2$  & $3.04 \times 10^{3}$ & $7.0\times 10^{-5}$ & 0.18 & 0.042 & $2.58\times 10^{-8}$  \\
		\multirow{-3}{*}{ $K_g (\phi)$}& $p = 2/3$    & $8.06 \times 10^{3}$ & $6.0\times 10^{-5}$ & $0.126$ & 0.10 & $2.76 \times 10^{-8}$  \\ \hline
		 	& $p = 1$ &$ 3.73\times 10^{3}$ & $5.14\times 10^{-3}$ & 0.40 & 0.00043 & $1.005\times 10^{-8}$  \\
		& $p = 2$ & $2.8\times 10^{3}$ & $4.26\times 10^{-3} $& 0.21 & 0.134 & $9.892\times 10^{-9}$  \\
	\multirow{-3}{*}{$K_s (\phi)$}	& $p = 2/3$  & $4.34\times 10^{3}$ & $5.29\times 10^{-3} $& 0.40 & 0.098 & $9.95\times 10^{-9}$  \\
		\hline \hline
	\end{tabular}}
	\caption{Benchmark points and predictions of gravitational wave spectrum $\Omega_{\text{GW}}$ and PBH abundance $Y_{\text{PBH}}$ for models $p = 1$, $p = 2$ and $p = 2/3$ generated using Gaussian peak function $K_g (\phi)$ and the step function $K_s (\phi)$.}
	\label{tb3}
\end{table}
\begin{table}[!htb]
	\setlength\extrarowheight{5pt}
	\centering
	\resizebox{\textwidth}{!}{
		\begin{tabular}{g|c|c|c|c|c|c } 
			\hline \hline \rowcolor{Gray}
			Model  & Experiment & $h$ & $w$ & $\phi_p$ & $Y_{\text{pbh}}^{\text{peak}}$ & $\Omega_{\text{GW}}^{\text{peak}}$ \\ 
			\hline
			& NG & $9.17 \times 10^{9}$ & $1.\times10^{-12}$ & 0.725 & $3.15\times 10^{-45}$ & $1.37\times10^{-9}$ \\
			& TL & $10.068\times10^{9}$ & $1.0\times10^{-12}$ & 0.42 & 0.089 & $1.26\times10^{-8}$\\
			& ET & $10.39\times10^{9}$ & $1.0\times10^{-12} $& 0.23 & 0.99 & $7.91\times 10^{-9}$\\
			\multirow{-4}{*}{$p = 1$}	& SE & $9.577\times 10^{9}$ & $1.0\times 10^{-12}$ & 0.534 & 0.19 & $2.97\times 10^{-8}$\\ \hline
			& NG & $6.06\times 10^{8}$ & $1.35\times 10^{-11}$ & 0.34 & $4.4\times 10^{-45}$ & $1.02\times 10^{-9}$ \\
			& TL & $5.008\times 10^{8}$ & $1.2\times 10^{-11}$ & 0.22 & 0.112 & $5.70\times 10^{-9}$\\
			& ET & $4.006\times 10^{8}$ & $1.0\times 10^{-11}$ & 0.146 & 0.87 & $1.01\times 10^{-8}$\\
			\multirow{-4}{*}{$p = 2$} & SE & $5.937\times 10^{8}$ & $1.35\times 10^{-11}$ & 0.31 & 0.177 & $2.19\times 10^{-8}$\\	\hline 
			& NG  & $9.17 \times 10^{9}$ & $1.\times10^{-12}$ & 0.725 & $3.15\times 10^{-31}$ & $1.37\times10^{-9}$ \\
			& TL   & $10.686\times10^{9}$ & $1.0\times10^{-12}$ & 0.58 & 0.091 & $1.54\times10^{-8}$\\
			& ET   & $12.26\times10^{9}$ & $1.0\times10^{-12} $& 0.42 & 0.83 & $9.74\times 10^{-9}$\\
			\multirow{-4}{*}{$p = 2/3$} & SE  & $9.733\times 10^{9}$ & $1.0\times 10^{-12}$ & 0.69 & 0.26 & $2.97\times 10^{-8}$\\ \hline \hline
	\end{tabular}}
	\caption{Benchmark points and predictions of gravitational wave spectrum $\Omega_{\text{GW}}$ and PBH abundance $Y_{\text{PBH}}$ for models $p = 1$, $p = 2$ and $p = 2/3$ generated using polynomial peak function $K_{q=2} (\phi)$. The labels NG, TL, ET and SE corresponds to peaks location of GW fraction $\Omega_{\text{GW}}(f)$ curve in regions of experiments; NANOGrav (NG), TaiJi/Lisa (TL), Einstein Telescope (ET) and SKA/EPTA (SE), respectively.}
	\label{tb2}
\end{table}
where $w$ is the width and $h$ is the height of the peaks. These peak functions are plotted in Fig. \ref{Kphi} as a function of $\phi$. It can be seen that the polynomial peak functions $K_{q=1,2} (\phi)$ have wider base and  falls off slowly. The Gaussian peak function $K_g (\phi)$ falls off exponentially with a narrow base whereas, the step peak function $K_s (\phi)$ has a flat top with the width exactly equal to $w$. Due to these peak functions, the first slow roll parameter $\epsilon$ becomes very small, $\mathcal{O} (\sim 10^{-10})$. This leads to ultra slow-roll inflation; the inflaton field gets trapped in this local minimum for $15$ to $40$ $e$-folds as shown in Fig. \ref{EEP}, where $\epsilon$, $\vert \eta \vert$ and $\phi$ are plotted against the number of $e$-folds $n = \log (a)$ for Gaussian peak function $K_g(\phi)$. Because of inverse relation between the power spectrum $P_\zeta$ and $\epsilon$, the sudden fall in the value of $\epsilon$ enhances the scalar power spectrum $P_\zeta$ by seven order of magnitude. This enhancement in power spectra can be realized in string theory inspired inflationary models due to large parametric space and the possibility of colliding branes \cite{Jennifer}. 

\begin{figure}[t]
	\centering \includegraphics[width=7.55cm]{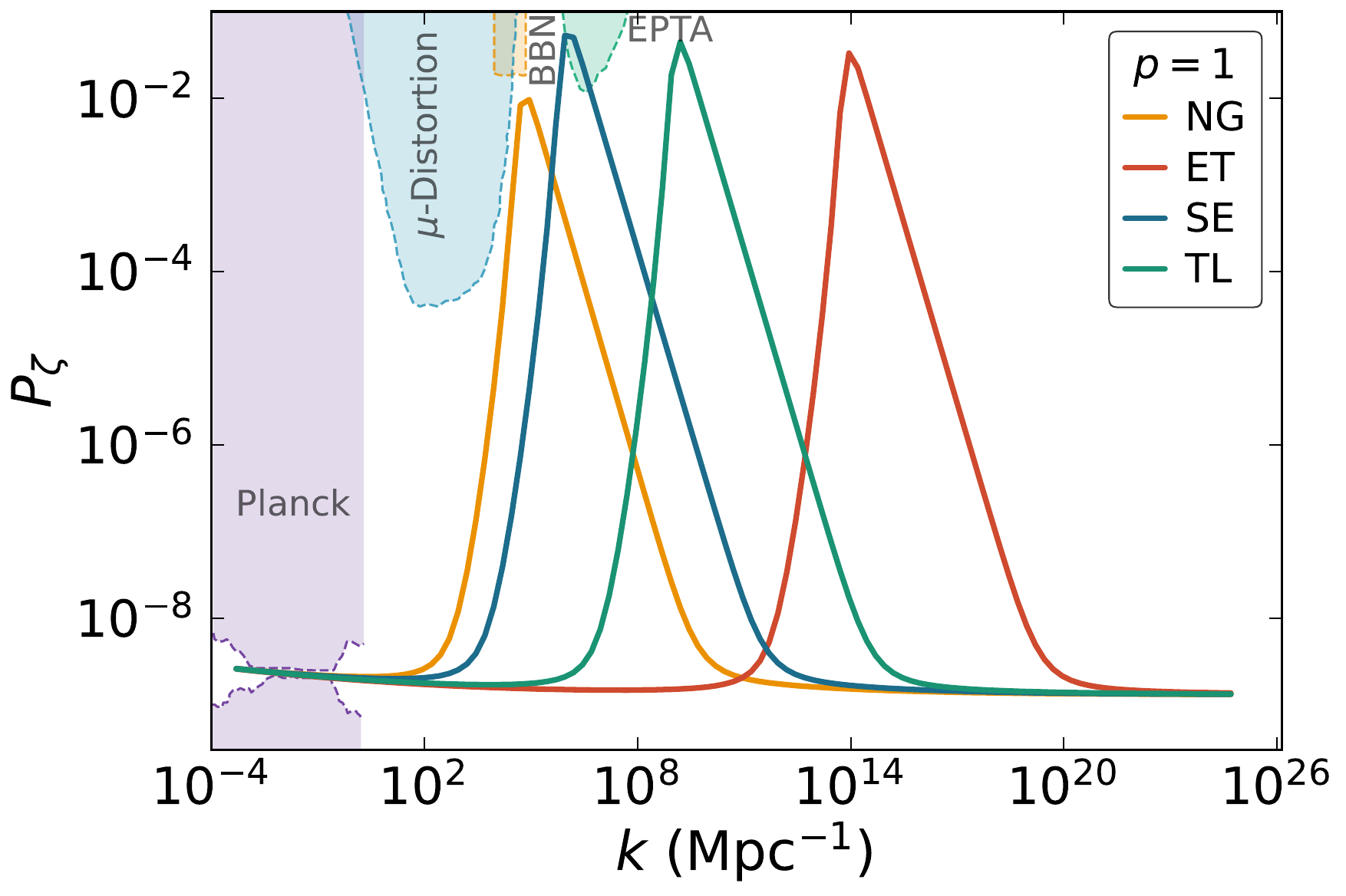}
	\centering \includegraphics[width=7.55cm]{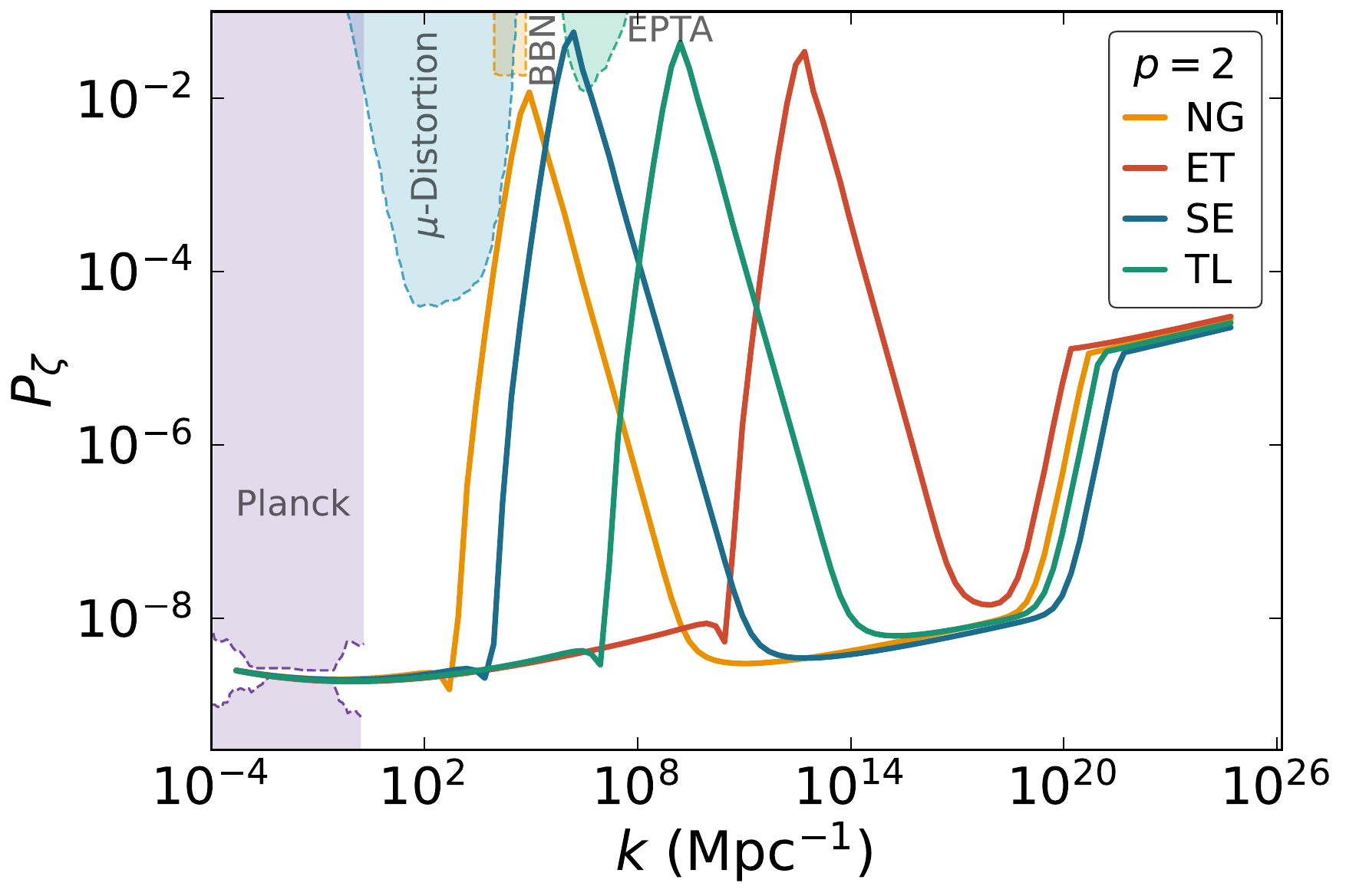}
	\centering \includegraphics[width=7.55cm]{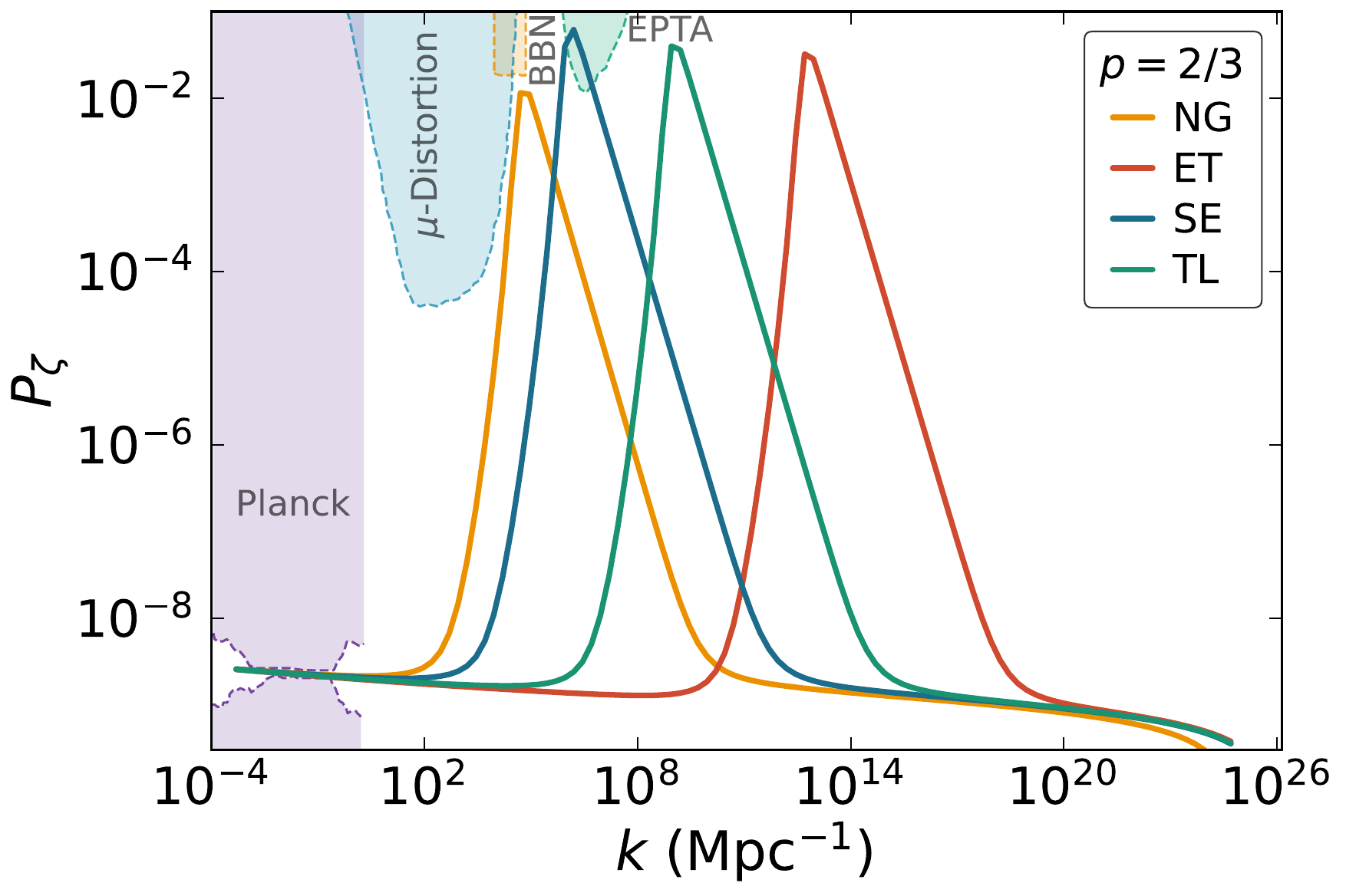}
	\caption{The scalar power spectrum $P_\zeta(k)$ generated using the polynomial peak function $K_{q}$ with $q=2$. The panels are drawn using parameter sets for the three models; $p = 1$ (upper left), $p = 2$ (upper right) and $p = 2/3$ (bottom), whereas the peaks in each panel correspond to parameter sets; NANOGrav (NG), TaiJi/Lisa (TL), Einstein Telescope (ET) and SKA/EPTA (SE), as listed in Table \ref{tb2}.}
	\label{fig:Pks_kq}
\end{figure}

We will evaluate power spectrum $P_\zeta(k)$, energy spectrum of induced gravitational wave $\Omega_{\text{GW}}(f)$ and PBH abundance $Y_{\text{PBH}}$ using  peak functions \eqref{kg} - \eqref{ks}. We ensure that the total number of $e$-folds vary between $56$ to $64$, enough to solve the horizon problem. The benchmark points and predictions of  gravitational waves spectrum $\Omega_{\text{GW}}$ and PBH abundance $Y_{\text{PBH}}$ for the hybrid inflation model with $p = 1$, $p = 2$ and $p = 2/3$ are listed in Table \ref{tb3} and \ref{tb2} for the peak functions $K_g (\phi)$, $K_s (\phi)$ and $K_{q=2} (\phi)$ along with the parameters $h$, $w$ and $\phi_p$. These parameters are chosen to enhance the power spectrum $\mathcal{O} (10^{-2})$ at small scales. 
\begin{figure}[t]
	\centering \includegraphics[width=7.55cm]{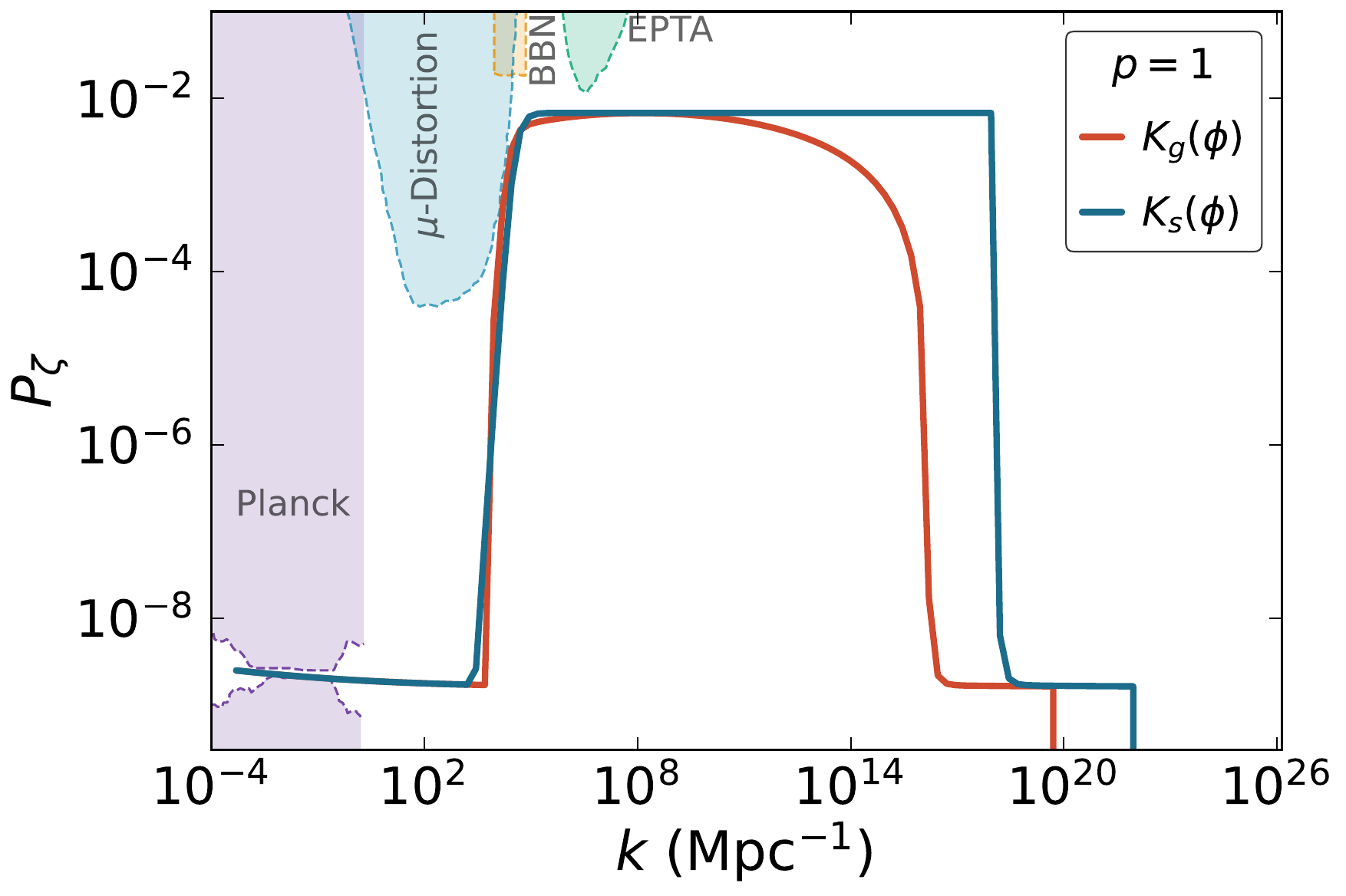}
	\centering \includegraphics[width=7.55cm]{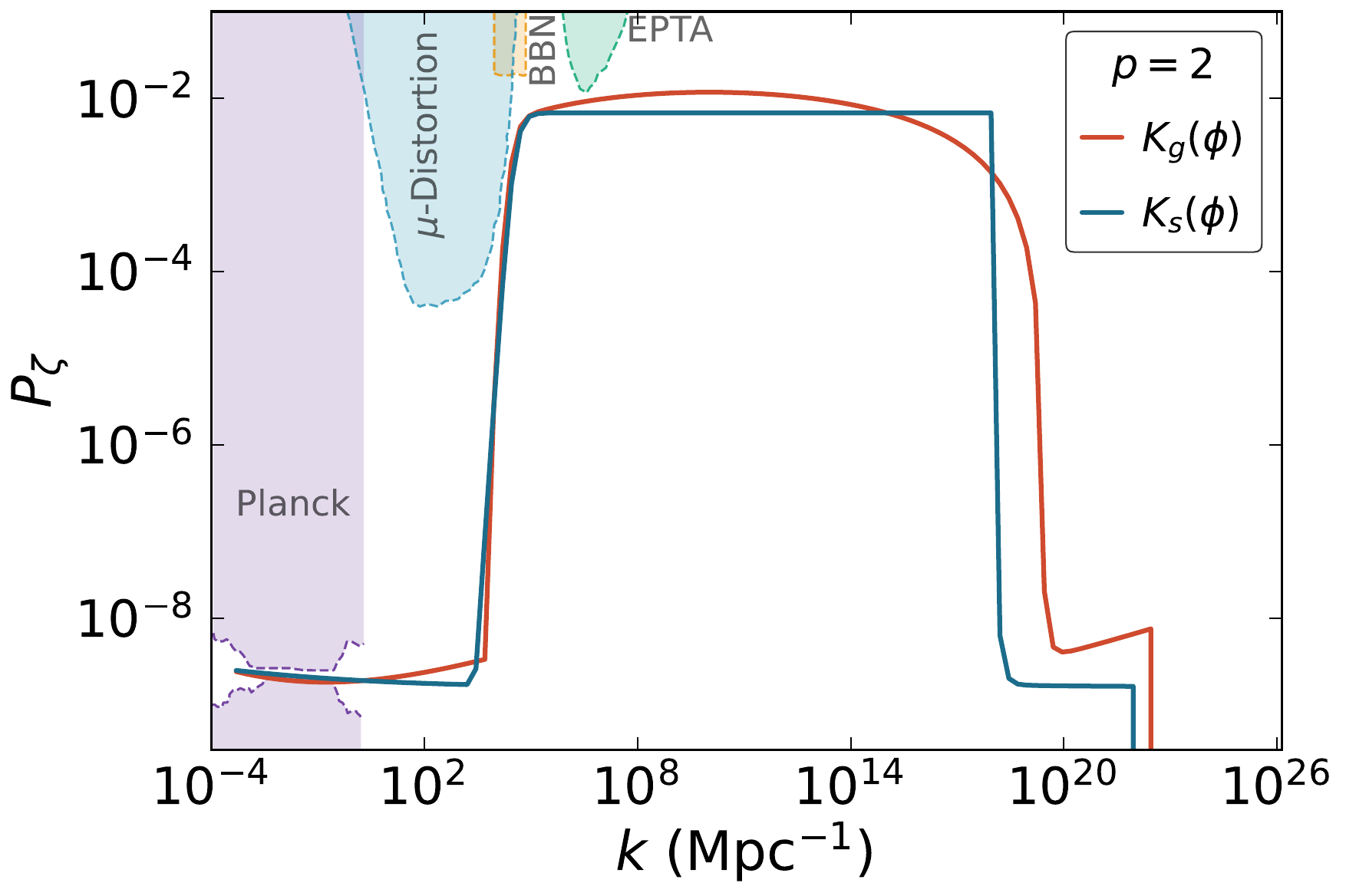}
	\centering \includegraphics[width=7.55cm]{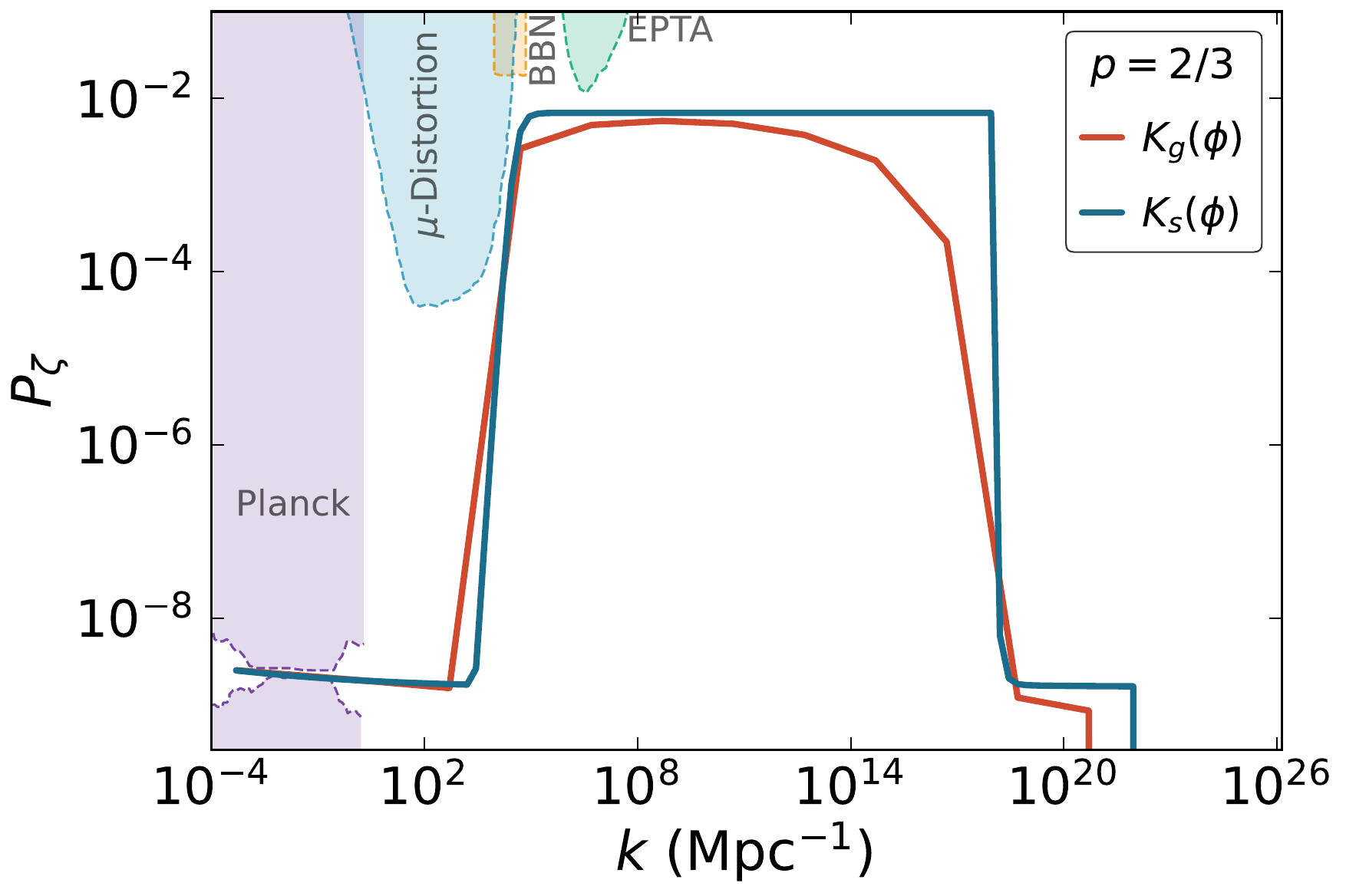}
\caption{The scalar power spectrum $P_\zeta(k)$ generated using the Gaussian $K_g(\phi)$ and step function $K_{s}(\phi)$. The panels are drawn using parameter sets for the three models; $p = 1$ (upper left), $p = 2$ (upper right) and $p = 2/3$ (bottom) as listed in Table \ref{tb3}. These functions generate broad power spectrum which appears extended and flat.}
\label{fig:Pks_ks_kg}
\end{figure}
The results for scalar power spectrum $P_\zeta(k)$ are shown in Figs. \ref{fig:Pks_ks_kg} and \ref{fig:Pks_kq} for our hybrid inflation model described by the scalar potential in Eq. \eqref{Vphi}. The curves in Fig. \ref{fig:Pks_ks_kg} are generated using the Gaussian peak function $K_g(\phi)$ and step function $K_{s}(\phi)$, defined in Eqs. \eqref{kg} and \eqref{ks}. The curves in Fig. \ref{fig:Pks_kq} are generated using the polynomial peak function $K_{q}(\phi)$ (defined in Eq. \eqref{kq}) with $q=2$, for the models $p = 1$, $p = 2$ and $p=2/3$. The four peaks  in each panel correspond to the parameter sets; NG, TL, ET and SE, listed in Table \ref{tb2}. 

It can be seen that at large scales $(0.05 \, \text{Mpc}^{-1})$, the power spectrum is of the order of $\mathcal{O}(10^{-9})$, compatible with CMB constraints. At small scales ($k > 10^{2} \, \text{Mpc}^{-1}$) however, the power spectrum is enhanced to the order $\mathcal{O} (10^{-2})$, large enough to produce PBHs after the horizon re-entry as discussed in the next section. Moreover, the Gaussian and step functions generate broad power spectrum which appears extended and flat, whereas the polynomial function generates peaked and narrow power spectrum. The broad power spectrum has important implications for GWs induced by the curvature perturbations responsible for PBH formation, as discussed in Sec. \ref{sec_4}. Note that the models satisfy the constraints from CMB $\mu$-distortion, big bang nucleosynthesis (BBN) and pulsar timing array (PTA) observations, except for the parameter set SE for polynomial peak function. 

\section{PBH Abundance} \label{sec_3}
%When the primordial curvature perturbation re-enters the horizon during radiation dominated era, it may gravitationally collapse to form PBHs. 

The gravitational collapse of primordial curvature perturbation upon horizon re-entry during radiation dominated era may yield PBHs. The PBH mass is equal to $\gamma M_{\mathrm{hor}}$, where $M_{\mathrm{hor}}$ is the horizon mass and we take $\gamma= 0.2$ \cite{Carr:1975qj}. The current fractional energy density of PBHs with mass $M$ to DM is \cite{Carr:2016drx}
\begin{equation}
Y_{\text{PBH}}(M)=\frac{\beta(M)}{3.94\times10^{-9}}\left(\frac{\gamma}{0.2}\right)^{1/2}
\left(\frac{g_*}{10.75}\right)^{-1/4}  \left(\frac{0.12}{\Omega_{\text{DM}}h^2}\right)
\left(\frac{M}{M_\odot}\right)^{-1/2},
\label{fpbheq1}
\end{equation}
where $M_{\odot}$ is the solar mass, $g_*$ is the effective degrees of freedom at the formation time
\begin{equation}
	g_* = \left\{
	\begin{array}{cc}
		10.75 & ~~~ 0.5 ~ \text{MeV} < T < 300 ~ \text{GeV} \\
		107.5 & T > 300 ~ \text{GeV}
	\end{array}\right. ,
\end{equation} 
and $\Omega_{\text{DM}}$ is the current energy density of the dark matter. The fractional energy density of PBHs at the formation is given by \cite{ Ozsoy:2018flq,Tada:2019amh}
\begin{equation}
\label{eq:beta}
\beta(M)\approx\sqrt{\frac{2}{\pi}}\frac{\sigma(M)}{\delta_c}
\exp\left(-\frac{\delta_c^2}{2\sigma^2(M)}\right),
\end{equation}
\begin{figure}[t]
	\centering \includegraphics[width=7.55cm]{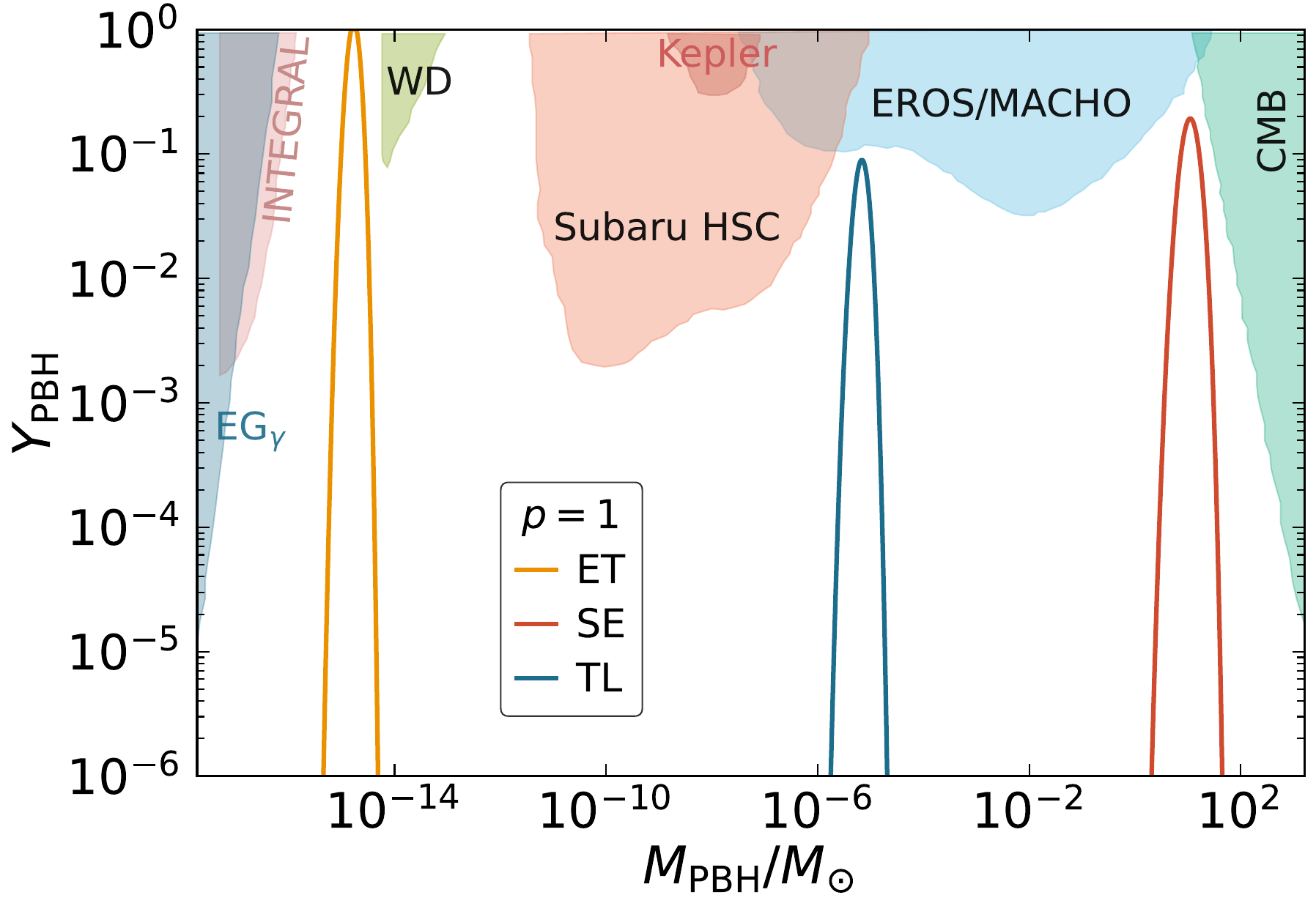}  
	\centering \includegraphics[width=7.55cm]{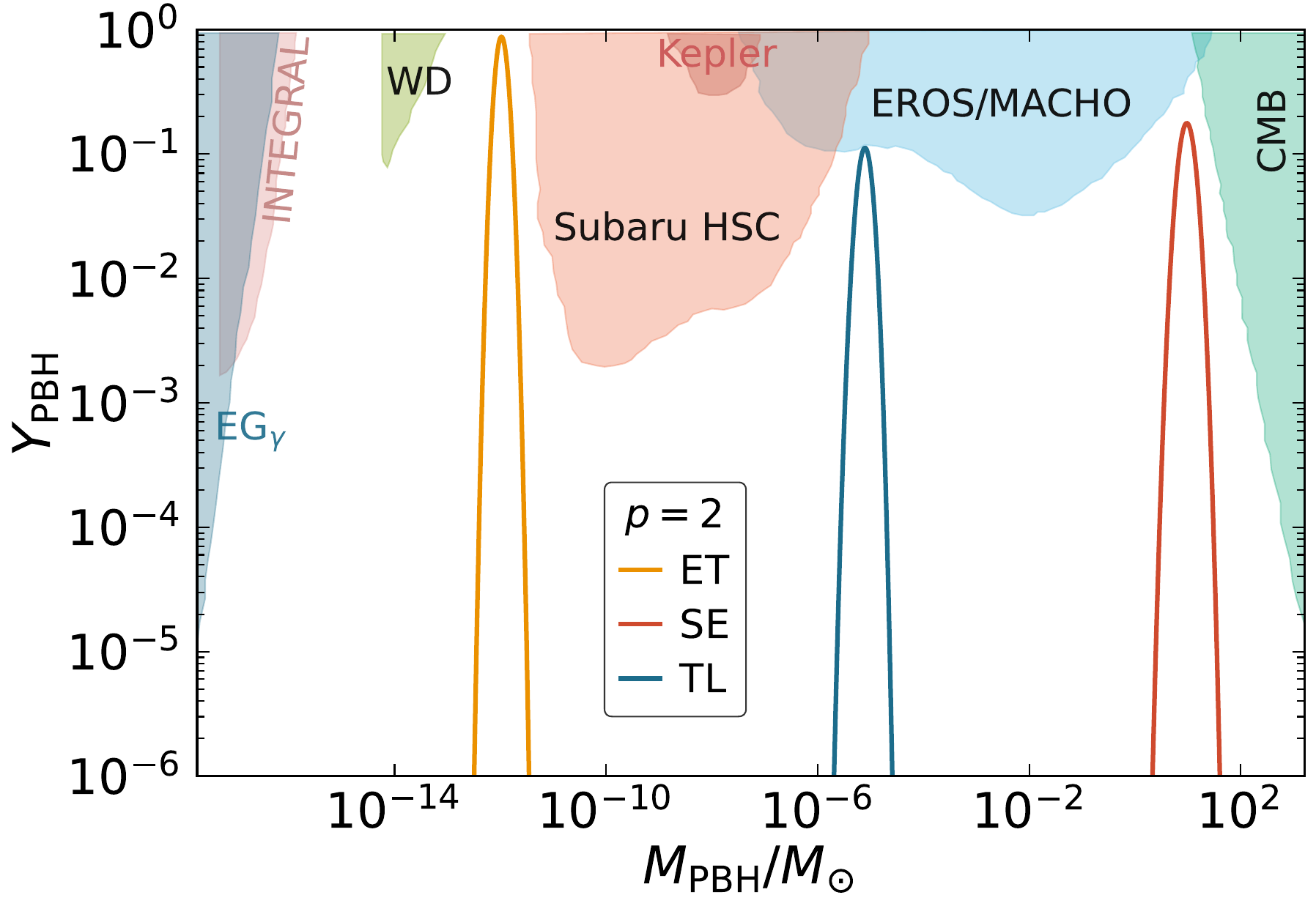}
	\centering \includegraphics[width=7.55cm]{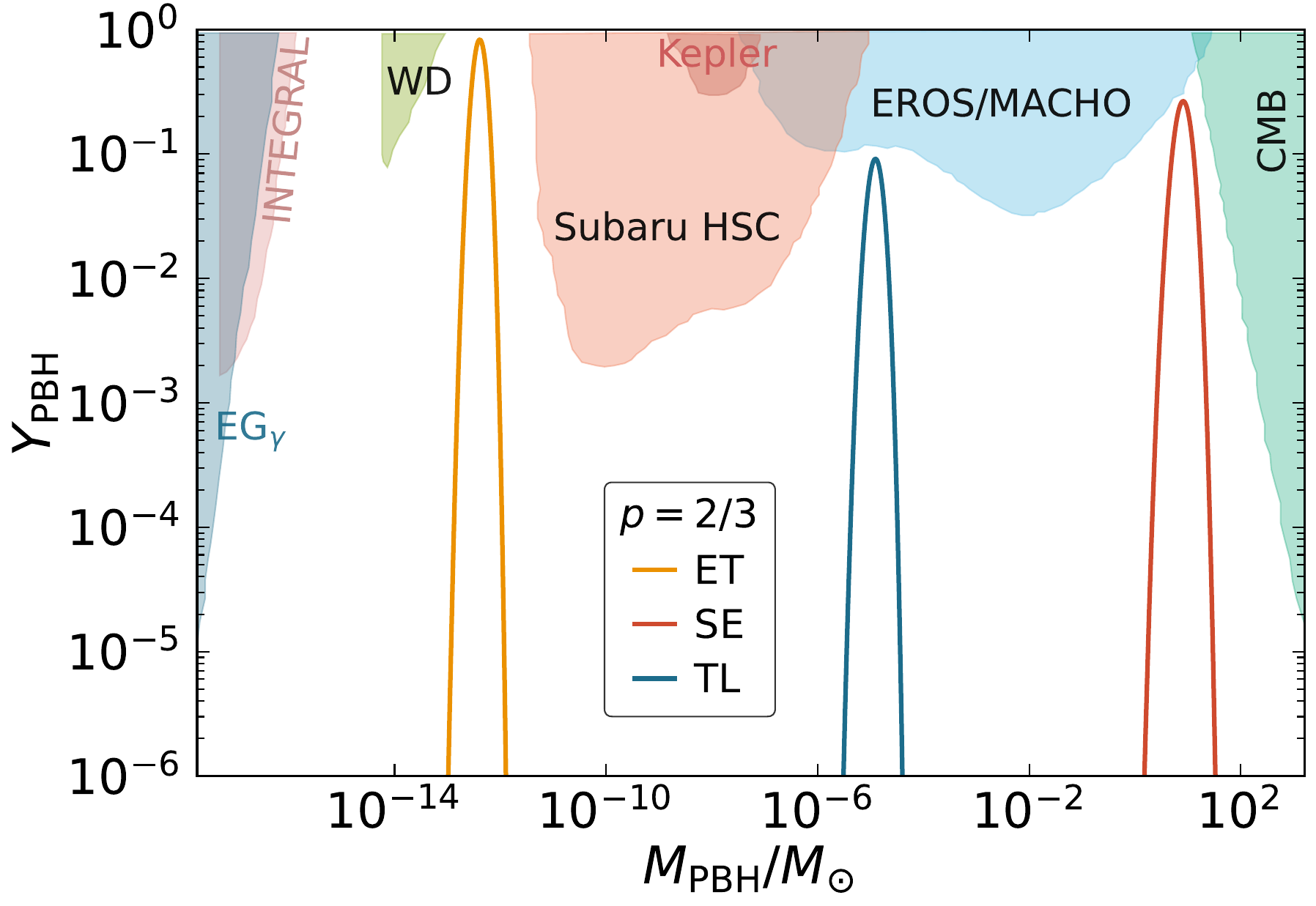}
	\caption{Primordial Black Hole (PBH) abundance $Y_{\text{PBH}}$ for polynomial peak function $K_{q} (\phi)$ with $q=2$. The three panels are drawn for the models $p=1$ (top left), $p=2$ (top right) and $p=2/3$ (bottom). The peaks in each panel correspond to the parameter sets; ET (Einstein Telescope), TL (TaiJi/Lisa) and SE (SKA/EPTA) from Table \ref{tb2}. The shaded regions represent the observational constraints on the PBH abundance from various experiments.}
	\label{Ypbh}
\end{figure} 
where $\delta_c$ denotes the critical density perturbation for the PBH formation and $\sigma(k)$ being the mass variance associated with the PBH mass $M(k)$ smoothing on the co-moving horizon length $k^{-1}=1/(aH)$, given by \cite{Ozsoy:2018flq}
\begin{equation}
\label{sigmaeq1}
\sigma^2(k)=\left(\frac{4}{9}\right)^2\int \frac{dq}{q} W^2(q/k)(q/k)^4P_{\zeta}(q),
\end{equation}
with the Gaussian window function $W(x)=\exp(-x^2/2)$. 
\begin{figure}[t]
	\centering \includegraphics[width=7.55cm]{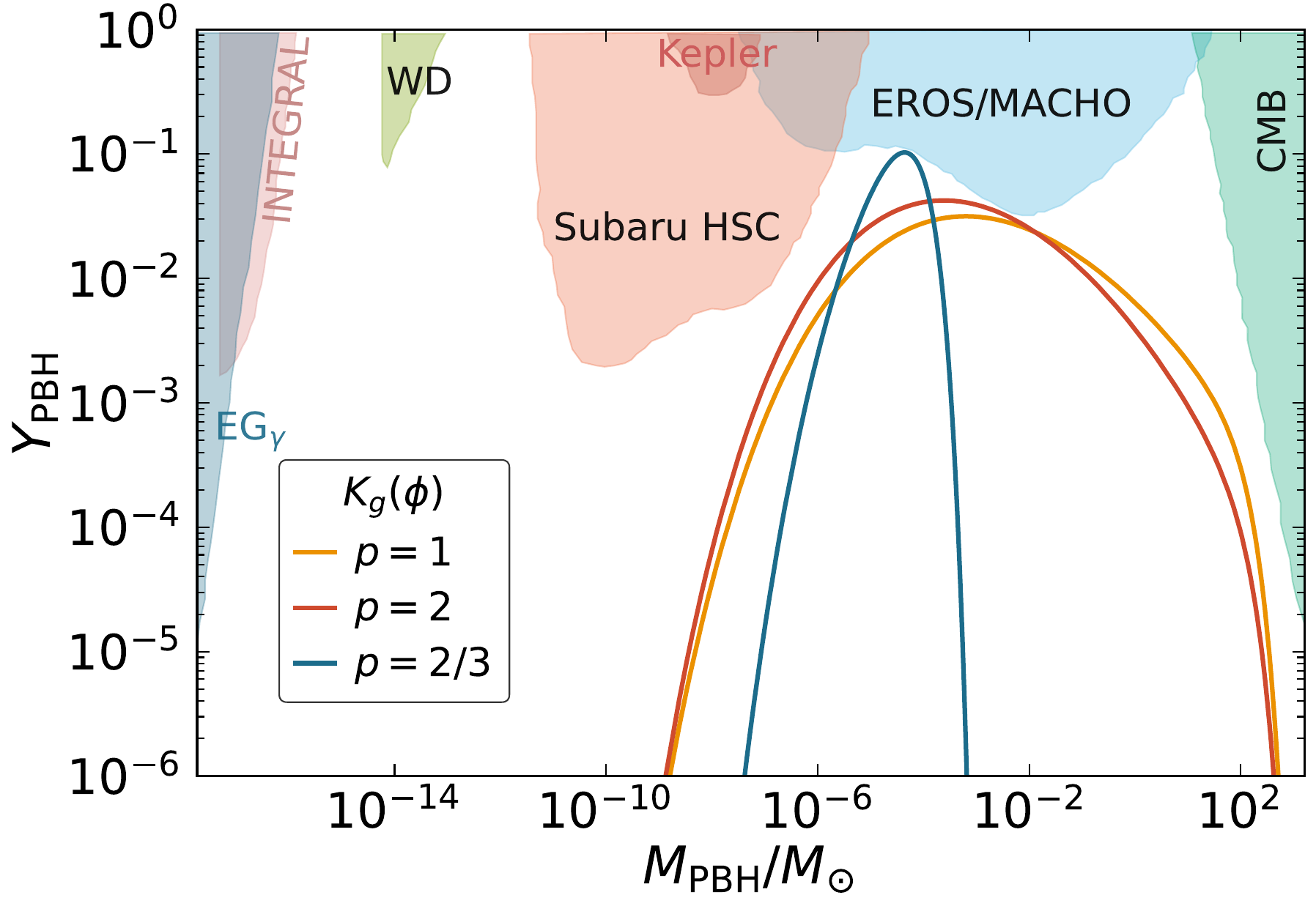}  
	\centering \includegraphics[width=7.55cm]{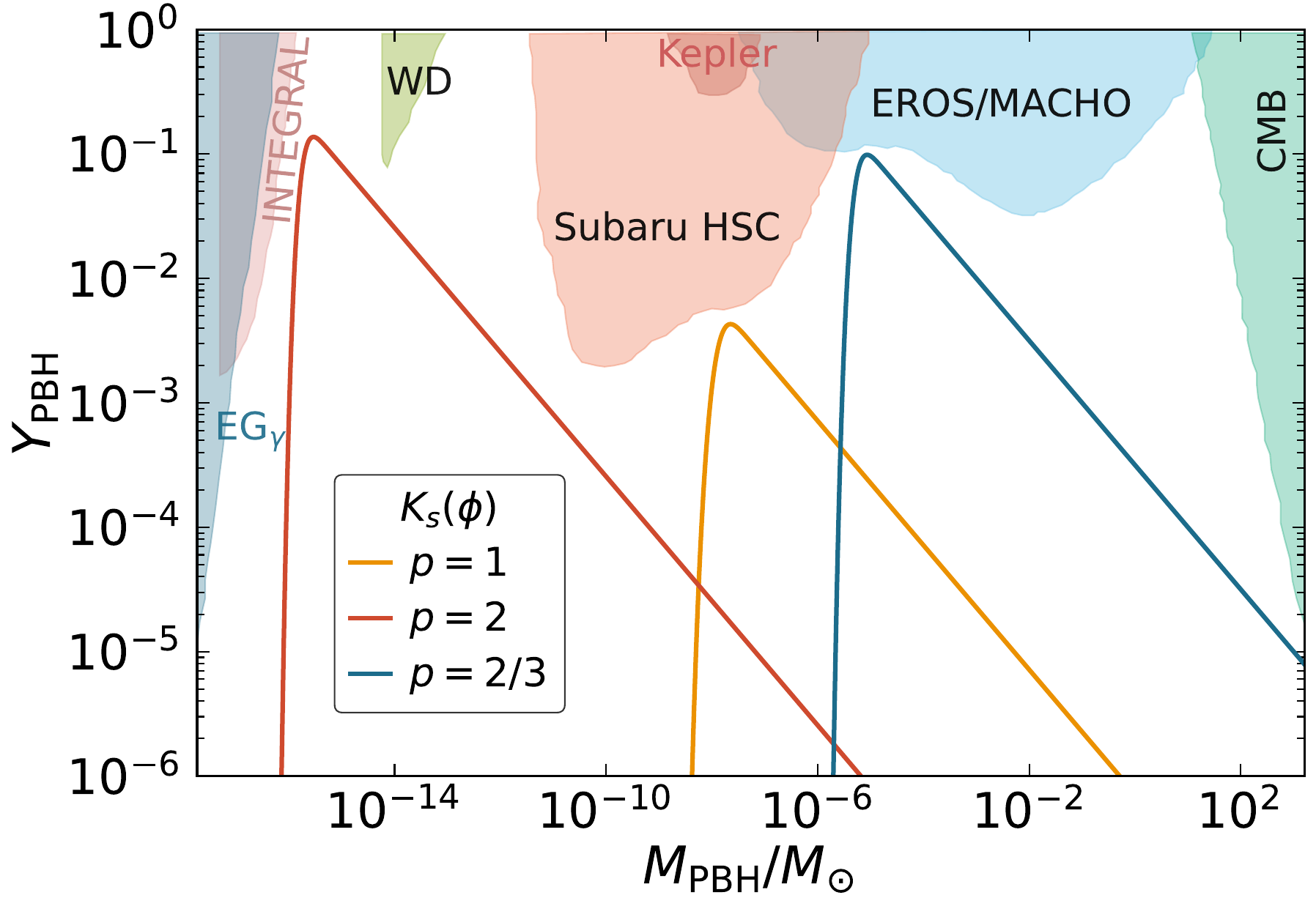}
	\caption{Primordial Black Hole (PBH) abundance $Y_{\text{PBH}}$ for Gaussian (left) and step (right) peak functions. The peaks correspond to the parameter sets for three models $p=1$, $p=2$ and $p=2/3$ from Table \ref{tb3}. The shaded regions represent the observational constraints on the PBH abundance from various experiments.}
	\label{Ypbh_2}
\end{figure} 
To calculate PBH abundance, we take the observational values; $\Omega_{\text{DM}}h^2=0.12$ \cite{Planck:2018vyg} and $\delta_c=0.4$ \cite{Tada:2019amh,Musco:2012au}. The relation between PBH mass $M$ and the scale $k$ is given by \cite{Ozsoy:2018flq}
\begin{equation}
\label{mkeq1}
M(k)=3.68\left(\frac{\gamma}{0.2}\right)\left(\frac{g_*}{10.75}\right)^{-1/6}
\left(\frac{k}{10^6\, \text{Mpc}^{-1}}\right)^{-2} M_{\odot}.
\end{equation}
Using the approximation that the power spectrum is scale invariant, we obtain
\begin{eqnarray}
	\sigma(k) &\simeq& (4/9)\sqrt{P_{\zeta}}, \\
	\beta(M) &\approx& \sqrt{\frac{2}{\pi}}\frac{\sqrt{P_{\zeta}}}{\mu_c}
	\exp\left(-\frac{\mu_c^2}{2P_{\zeta}}\right),
\end{eqnarray}
 where $\mu_c=9\delta_c/4$. Substituting the power spectrum obtained above into Eqs. \eqref{fpbheq1}, \eqref{eq:beta}, \eqref{sigmaeq1} and \eqref{mkeq1}, we obtain the PBH abundances as displayed in Fig. \ref{Ypbh} for polynomial peak function $K_{q} (\phi)$ with $q=2$ and in Fig. \ref{Ypbh_2} for Gaussian $K_g (\phi)$ and step peak function $K_s (\phi)$. The three panels in Fig. \ref{Ypbh} are drawn for the models $p=1$, $p=2$ and $p=2/3$ whereas, the peaks correspond to the parameter sets ET, TL and SE from Table \ref{tb2}. The shaded regions in the background represent the observational constraints on the PBH abundance from different experiments such as; accretion constraints by CMB, extragalactic gamma-rays by PBH evaporation (EG$_\gamma$), galactic center 511 keV gamma-ray line (INTEGRAL), white dwarf explosion (WD), microlensing events with Subaru HSC, the Kepler satellite and EROS/MACHO. For the polynomial peak function $K_q(\phi)$ with $q = 2$ the models $p = 1$, $p = 2$ and $p = 2/3$ produce PBHs with masses $M_{\text{PBH}} \simeq (10^{-15},\, 11, \, 10^{-5}) \, M_{\odot}$, $M_{\text{PBH}} \simeq (10^{-12},\, 10, \, 10^{-5}) \, M_{\odot}$, $M_{\text{PBH}} \simeq (10^{-13},\, 8, \, 10^{-5}) \, M_{\odot}$ and abundances $Y_{\text{PBH}}^{\text{peak}} \simeq (0.99, 0.19, 1.09)$, $Y_{\text{PBH}}^{\text{peak}} \simeq (0.87, 0.18, 0.11)$, $Y_{\text{PBH}}^{\text{peak}} \simeq (0.83, 0.26, 0.09)$ for parameter sets (ET, SE TL), respectively. Similarly, for the Gaussian peak function $K_g (\phi)$, the parameter sets for models $p = (1, 2, 2/3)$ produce PBHs with masses $M_{\text{PBH}} \simeq (10^{-3}, 10^{-4}, 10^{-5}) \, M_{\odot}$ and abundances $Y_{\text{PBH}}^{\text{peak}} \simeq (0.03, 0.04, 0.1)$, respectively. Finally, for the step peak function $K_s (\phi)$, the parameter sets for models $p = (1, 2, 2/3)$ produce PBHs with masses $M_{\text{PBH}} \simeq (10^{-8}, 10^{-16}, 10^{-5}) \, M_{\odot}$ and abundances $Y_{\text{PBH}}^{\text{peak}} \simeq (0.004, 0.14, 0.098)$, respectively. Note that there are no observational constraints on PBH abundances in these mass ranges and therefore, PBHs can constitute all of the dark matter (DM).

\section{Production of Secondary Gravitational Waves} \label{sec_4}
The production of primordial black holes (PBHs) due to large curvature or density perturbations can induce secondary GWs due to second order mode coupling. These Secondary Induced GW (SIGW) are gauge invariant \cite{DeLuca:2020agl} stochastic background waves that could be observed by future GW experiments.
\begin{figure}[t]
	\centering \includegraphics[width=7.55cm]{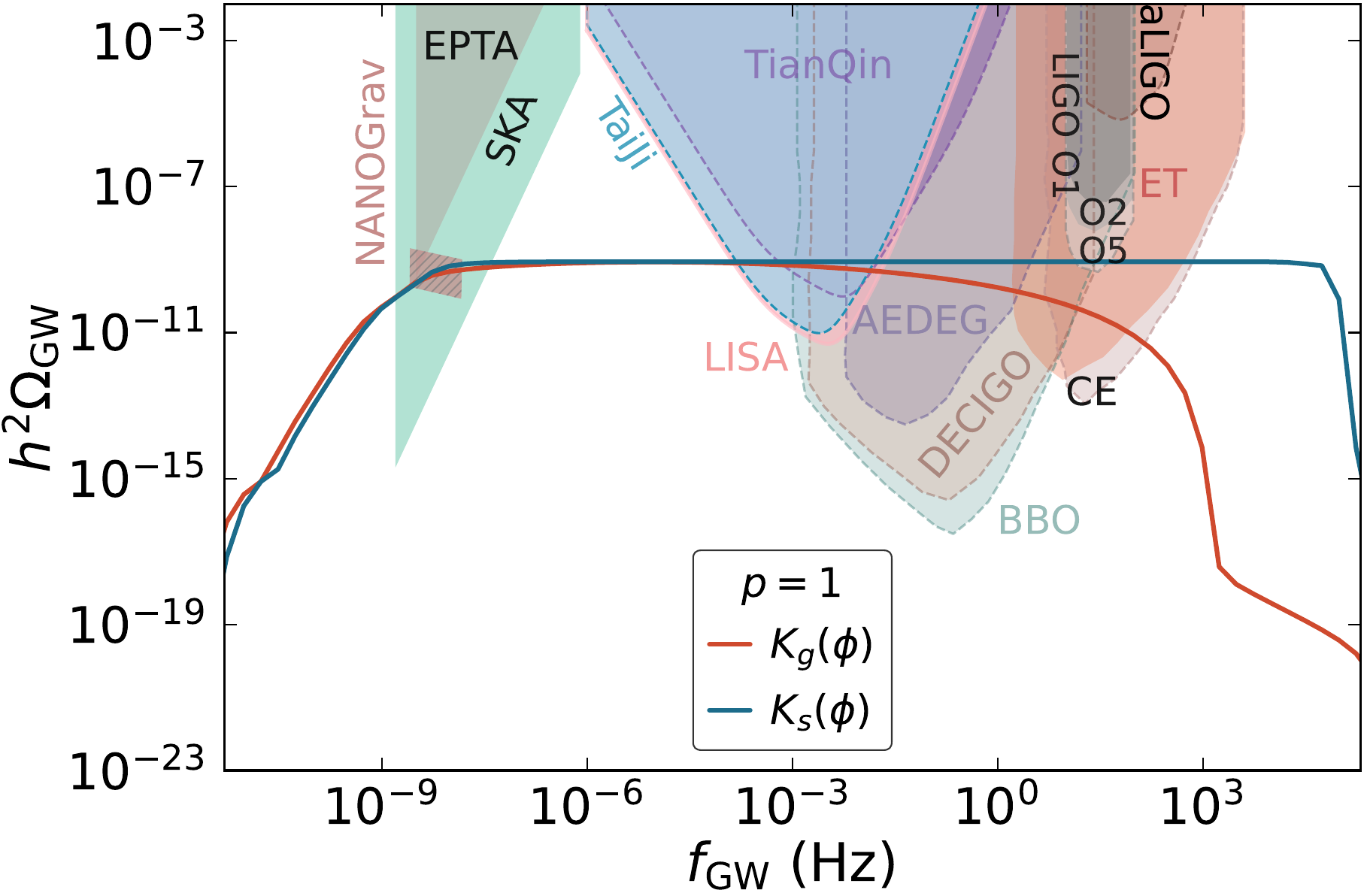}
	\centering \includegraphics[width=7.55cm]{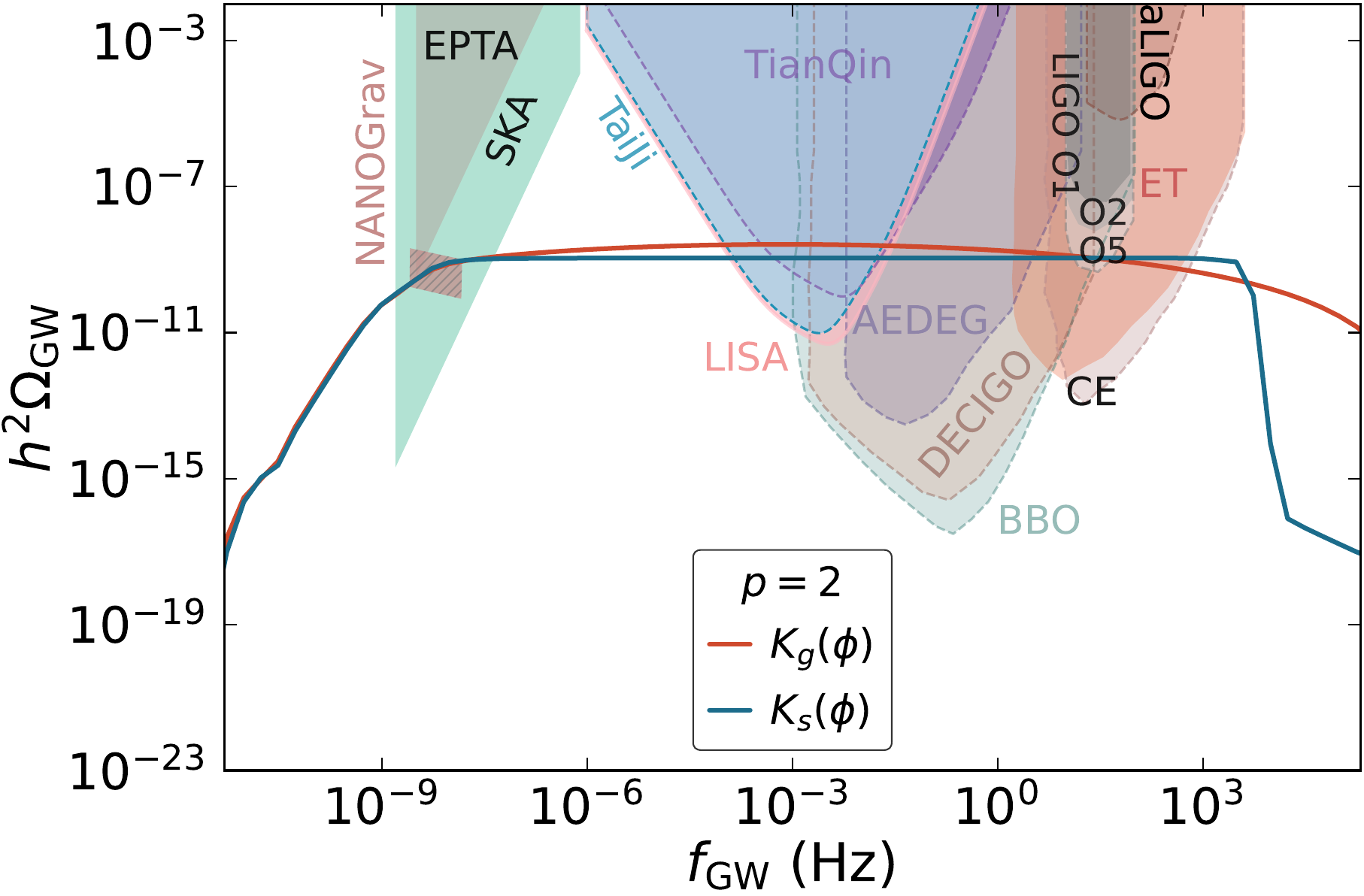}
	\centering \includegraphics[width=7.55cm]{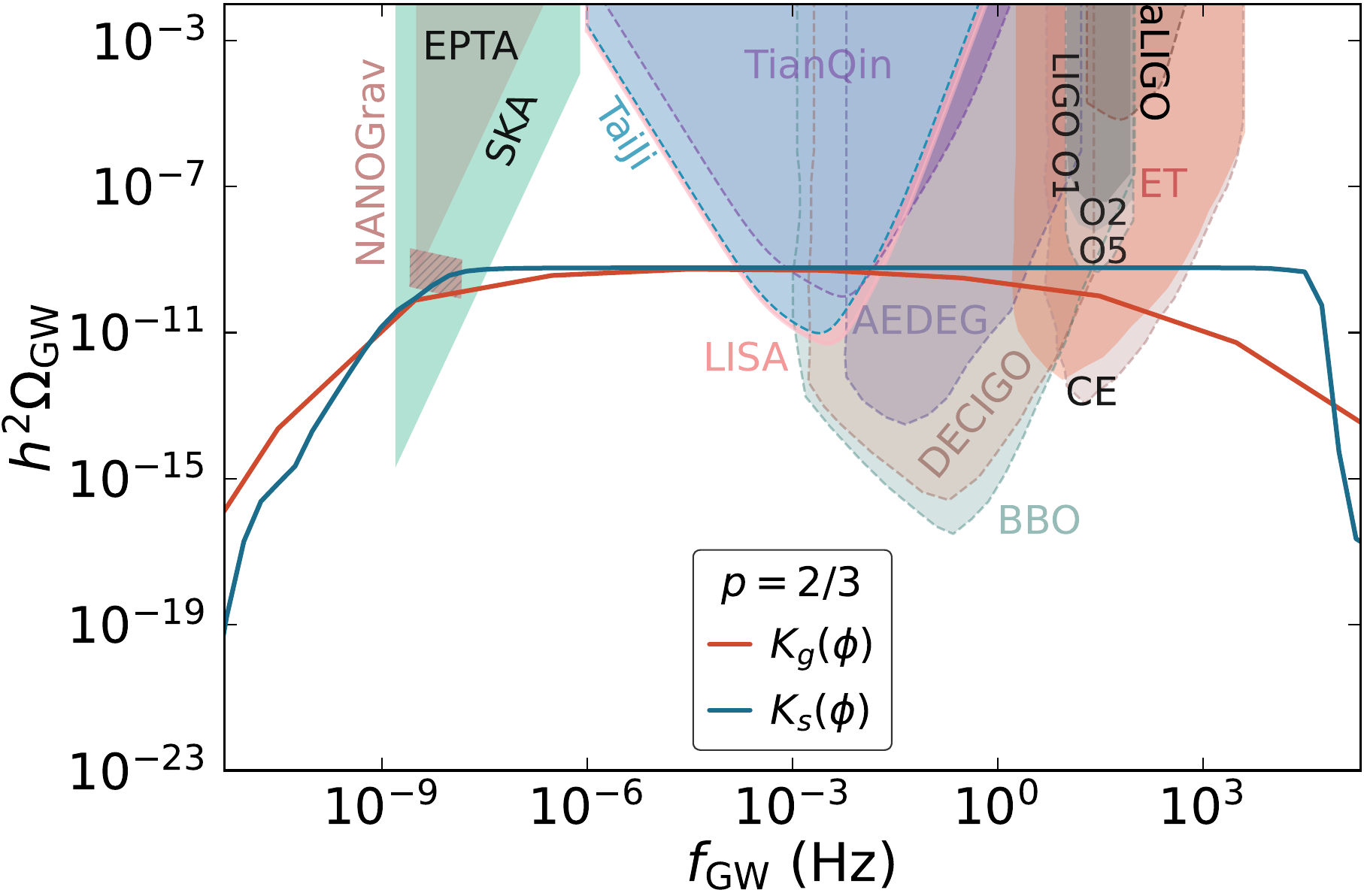}
	\caption{Gravitational waves spectrum as a function of the frequency of gravitational wave, generated for hybrid inflation model with the Gaussian $K_{g} (\phi)$ and step $K_{s} (\phi)$ functions. The three panels are drawn using parameter set for each model $p = 1$, $p = 2$ and $p = 2/3$ as listed in Table \ref{tb3}. The shaded regions in the background represent the sensitivity of current and future gravitational waves observatories.}
	\label{Ogw_gs}
\end{figure}
\begin{figure}[!htb]
	\centering \includegraphics[width=7.55cm]{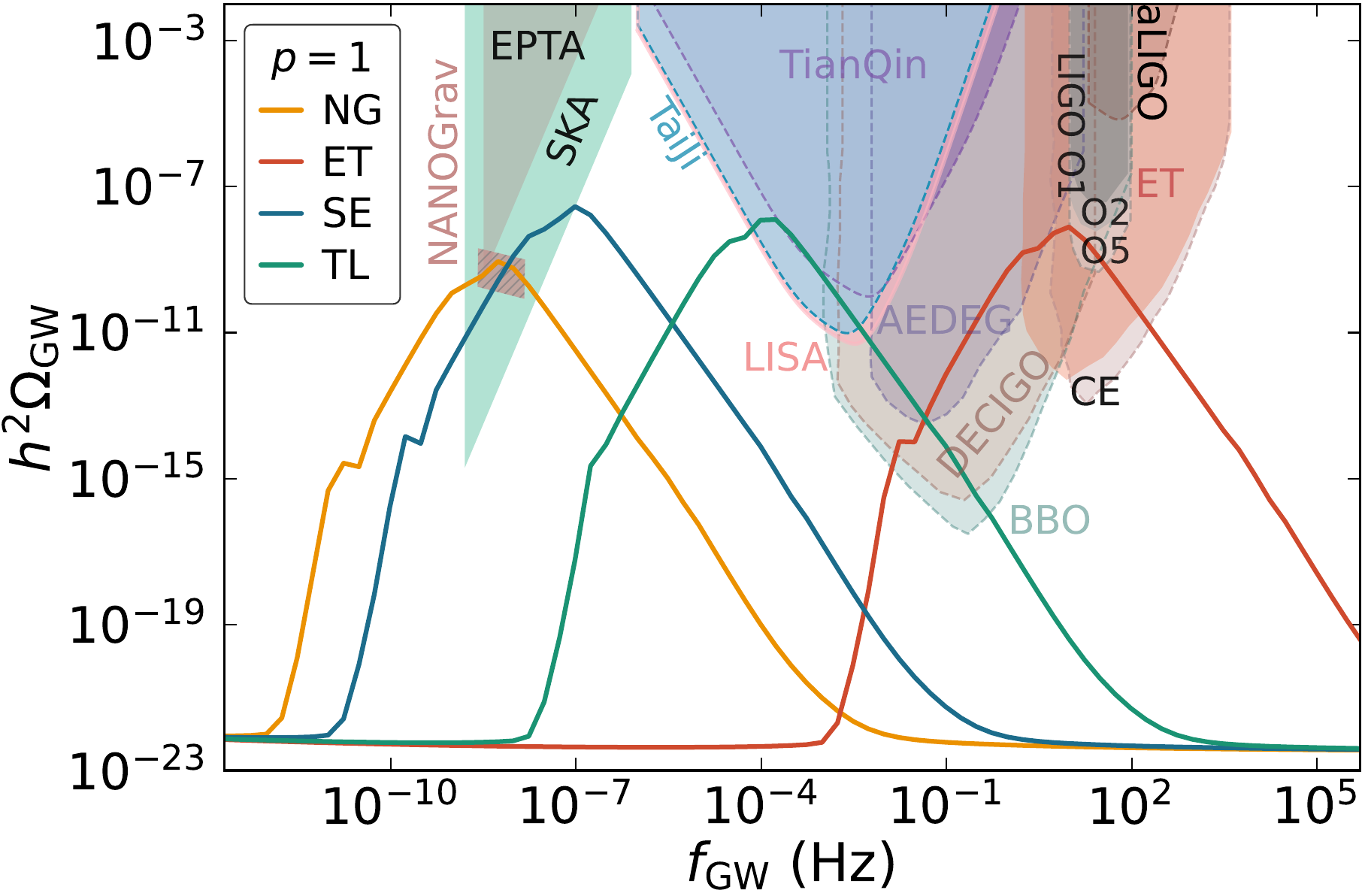}
	\centering \includegraphics[width=7.55cm]{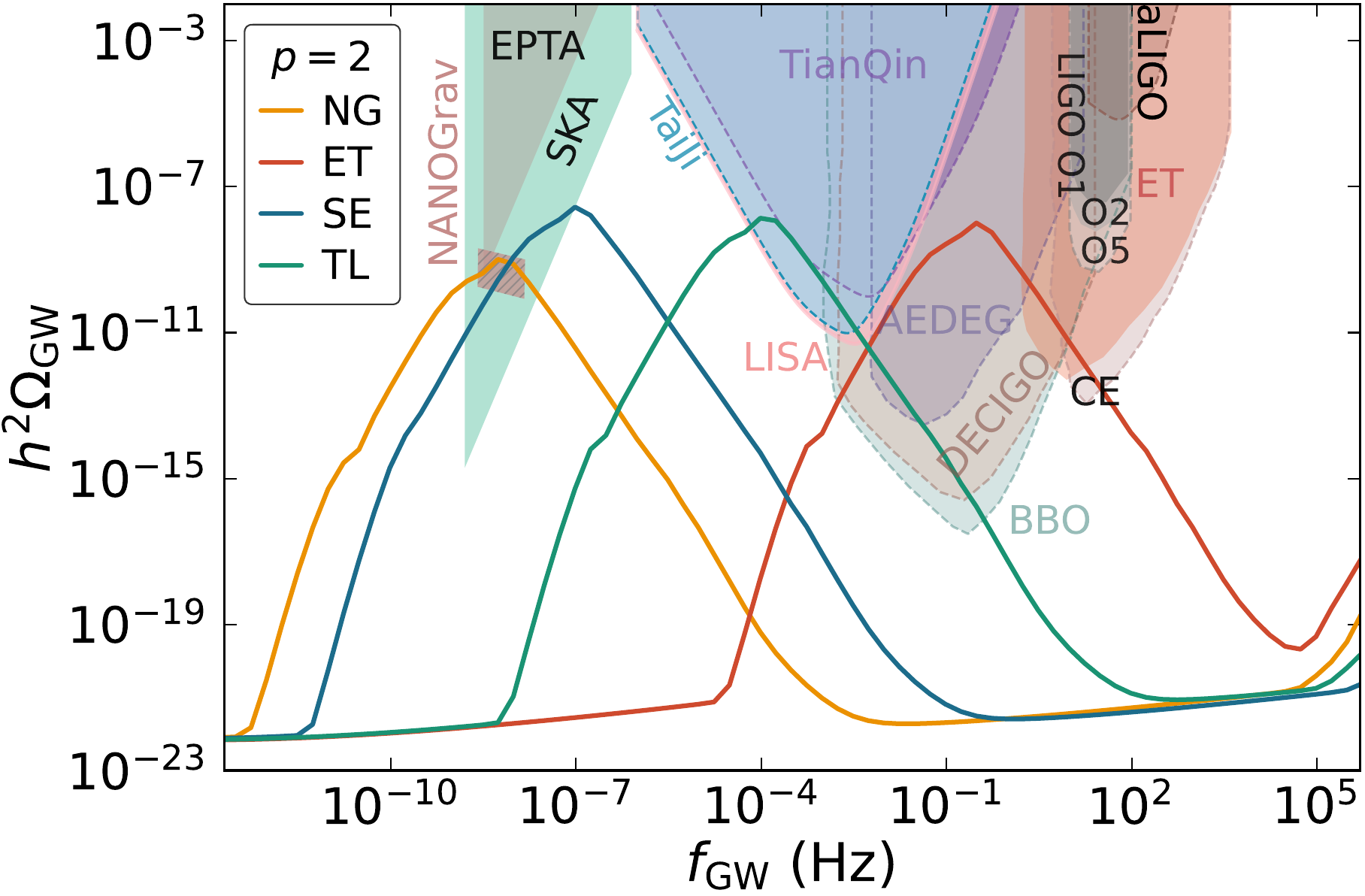}
	\centering \includegraphics[width=7.55cm]{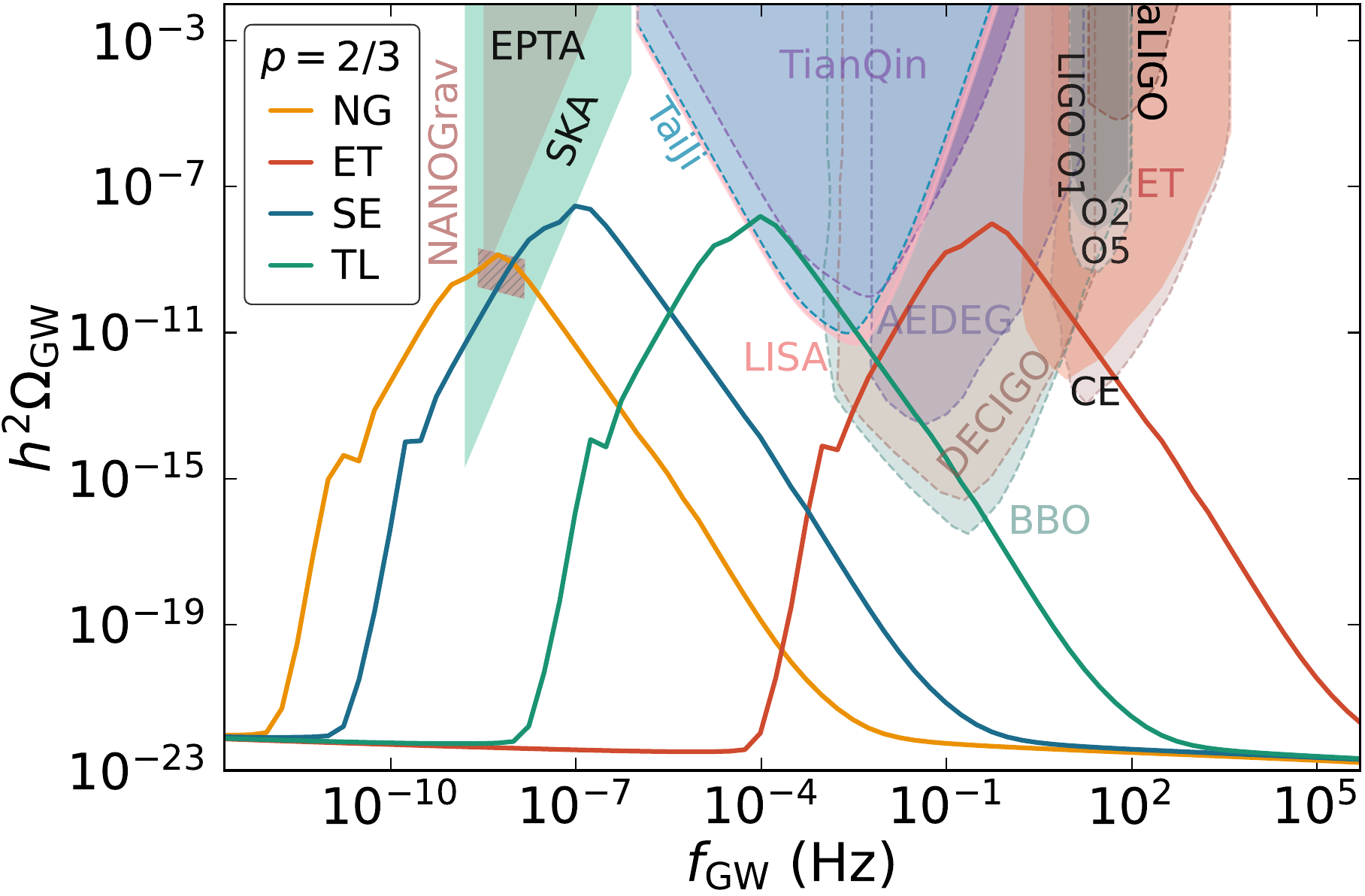}	    
	\caption{Gravitational waves spectrum as a function of the frequency of gravitational wave, generated for hybrid inflation model with the polynomial peak function $K_{q} (\phi)$, with $q=2$. The peaks correspond to the parameter sets; NANOGrav (NG), TaiJi/Lisa (TL), Einstein Telescope (ET) and SKA/EPTA (SE), as listed in Table \ref{tb2}. The shaded regions in the background represent the sensitivity of current and future gravitational waves observatories.}
	\label{Ogw_q}
\end{figure}
\begin{figure}[t]
	\centering \includegraphics[width=7.55cm]{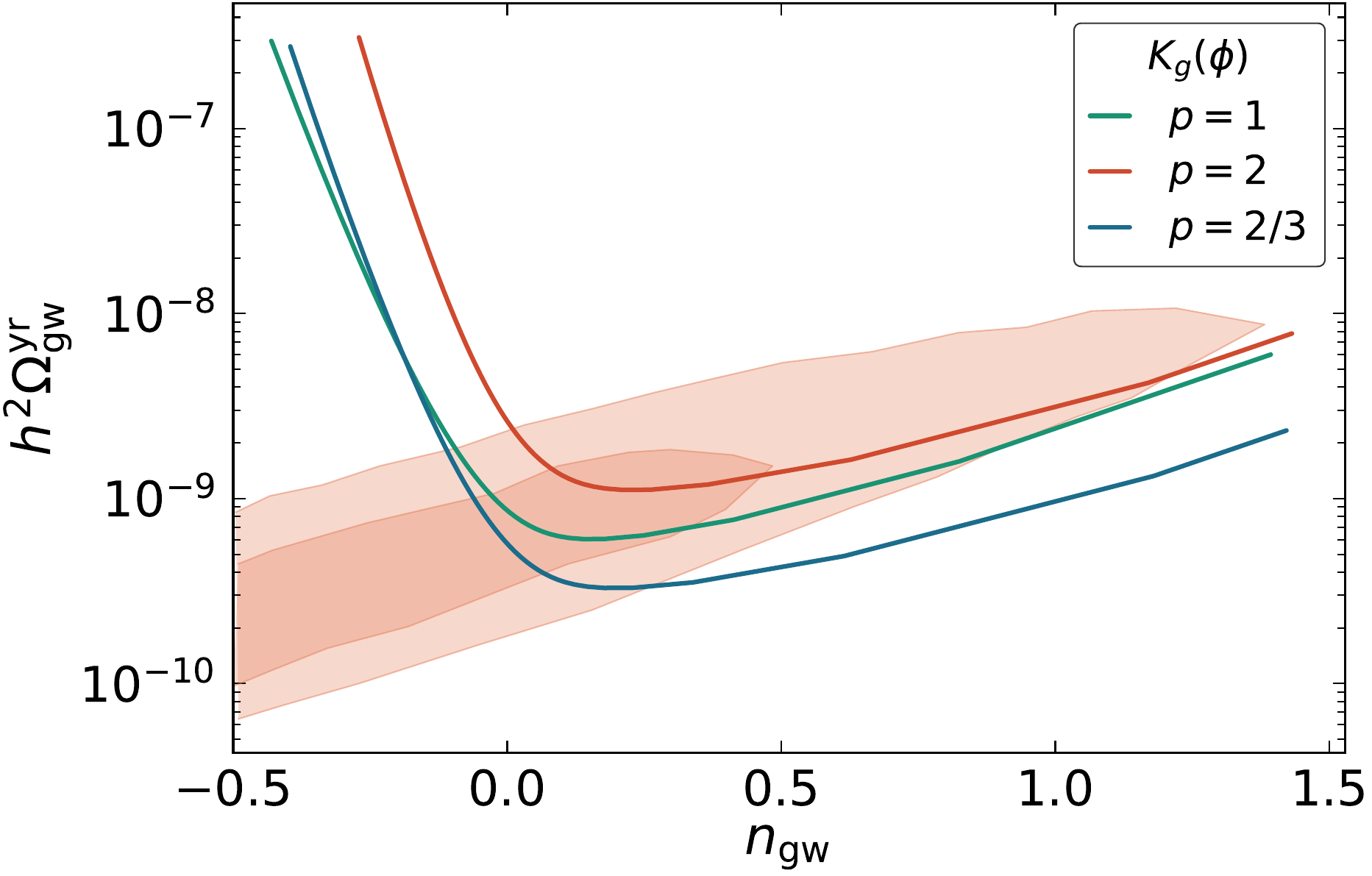}
	\centering \includegraphics[width=7.55cm]{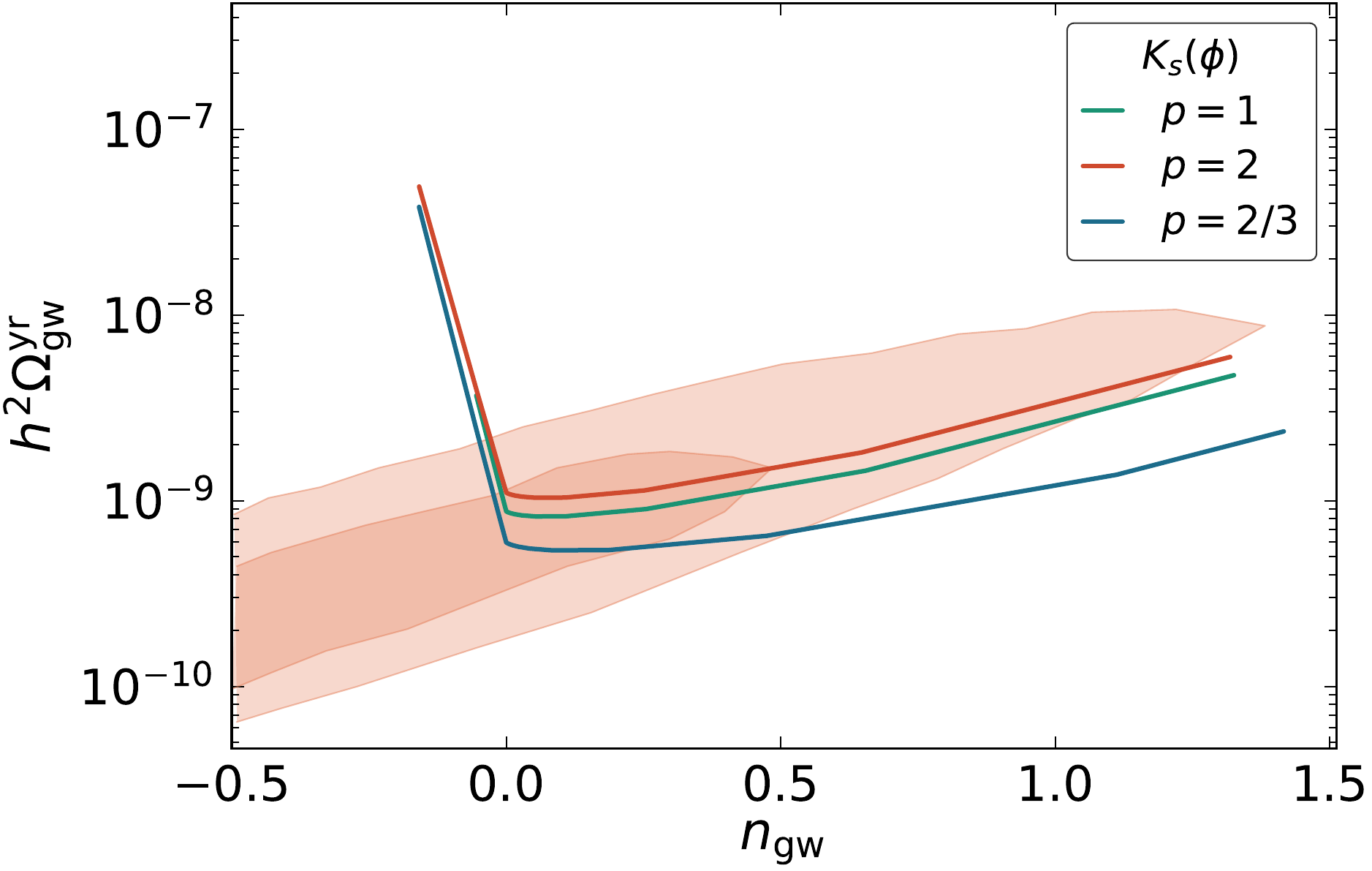}
	\caption{Gravitational wave signals from hybrid inflation for the models $p = 2$, $p = 1$ and $p = 2/3$ compared to the NANOGrav observations. The dark and light shaded regions represent 1- and 2-$\sigma$ bounds reported by NANOGrav \cite{Arzoumanian:2020vkk}. The left panel corresponds to the Gaussian peak function $K_{g} (\phi)$, whereas the right panel corresponds to the step function $K_s (\phi)$. }
	\label{YrGw}
\end{figure}

The tensor mode $h_k$ for SIGW is sourced by quadratic scalar perturbations function $S_k(\Phi_k)$ \cite{Ananda:2006af,Baumann:2007zm}
\begin{eqnarray}
h''_k&+&2\mathcal{H}h'_k+k^2k_k = 4S_k(\Phi_k),\\
S_k(\Phi_k)&=&\int \frac{d^3q}{(2\pi)^{\frac{3}{2}}}q^i q^j e_{ij}(k)\left( 2\Phi_q\Phi_{k-q}+\frac{4(\Phi'_q+\mathcal{H}\Phi_q)}{3(1+w)\mathcal{H}^2}\left(\Phi'_{k-q}+\mathcal{H}\Phi_{k-q} \right) \right),
\end{eqnarray}
where $\mathcal{H}=aH$, $w=p/\rho$, $e_{ij}(k)$ is polarization tensor and $\Phi_k$ is the gauge invariant Bardeen potential\cite{bardeen80}. During the radiation dominated era, these SIGWs decouple from their scalar part and plateau after horizon crossing. The energy density of these SIGWs today is given by \cite{DeLuca:2020agl,Terada}
\begin{eqnarray}
\Omega_{\text{GW}}(k)&=& 0.387 \frac{\Omega_r}{6} \left( \frac{g^4_{*,s}g_*^{-3}}{106.75} \right)^{-\frac{1}{3}} \nonumber \\
&\times& \int_{-1}^{1} dx \int_{1}^{\infty}dy~P_\zeta\left(k \frac{y-x}{2} \right) P_\zeta\left(k \frac{x+y}{2} \right) F(x,y)\label{OmgaGW} ,
\end{eqnarray}
where
\begin{eqnarray}
F(x,y)&=&\frac{288 (x^2 + y^2 - 6)^2 (x^2 - 1)^2 (y^2 - 1)^2}{(x - y)^8 (x + y)^8}  \nonumber \\
&\times&\Bigg[ \left( x^2 - y^2 + \frac{x^2 + y^2 - 6}{2} \log\left| \frac{y^2-3}{x^2-3} \right| \right)^2 \nonumber \\
 &+&  \frac{\pi^2}{4}(x^2 + y^2 - 6)^2 \theta(y-\sqrt{3}) \Bigg].
\end{eqnarray}
In the above expression, $\Omega_r=5.38\times 10^{-5}$ is the radiation energy density and $g_{*}$, $g_{*,s}$ are effective degrees of freedom at horizon crossing for each mode. Because of second order mode coupling, $\Omega_{\text{GW}}$ has quadratic dependence on scalar power spectrum $P_\zeta(k)$. 

The GW spectrum for our hybrid inflation model is shown in Fig. \ref{Ogw_gs} with Gaussian $K_g (\phi)$ and step peak function $K_s (\phi)$ and in Fig. \ref{Ogw_q} with the polynomial peak function $K_{q} (\phi)$, $q=2$. The peaks in these figures correspond to the parameter sets listed in Table \ref{tb3} for functions $K_g (\phi)$ and $K_s (\phi)$ and Table \ref{tb2} for the polynomial function $K_{q} (\phi)$. 

It is evident from Fig. \ref{Ogw_gs} that the Gaussian and step functions generate broad gravitational wave spectrum $\Omega_{\text{GW}} (k)$ in the frequency range $f_{\text{GW}} \simeq (10^{-9} - 10^{3})$ Hz which can be detected by current and future gravitational wave detectors. These include ground based detectors such as; Square Kilometre Array (SKA)\cite{Smits:2008cf}, North American Nanohertz Observatory for Gravitational Waves (NANOGrav)\cite{Arzoumanian:2020vkk}, Einstein Telescope (ET) \cite{Punturo:2010zz}, Cosmic Explorer (CE) \cite{LIGOScientific:2016wof}, Laser Interferometer Gravitational-Wave Observatory (LIGO)  O5 \cite{LIGOScientific:2019vic}, and spaced based detectors such as; Laser Interferometer Space Antenna (LISA)\cite{LISA:2017pwj}, Deci-hertz Interferometer Gravitational wave Observatory (DECIGO) \cite{Seto:2001qf}, Big Bang Observer (BBO) \cite{Corbin:2005ny}, TaiJi \cite{Hu:2017mde}, TianQin \cite{TianQin:2015yph} and Atomic Experiment for Dark Matter and Gravity Exploration in Space (AEDGE) \cite{AEDGE:2019nxb}. The sensitivity bounds of Advanced LIGO, European Pulsar Timing Array (EPTA) and LIGO O1/LIGO O2 are also included, although the peaks lie outside these bounds. Note that all the peaks in Fig. \ref{Ogw_gs} lie in the sensitivity bounds of NANOGrav and these SIGWs signals, associated with the formation of primordial black holes (PBHs), may be the source of stochastic process recently reported by NANOGrav analysis of 12.5 yrs of data \cite{Arzoumanian:2020vkk}.

The polynomial peak function $K_q (\phi)$, $q = 2$ generates narrow $\Omega_{\text{GW}} (k) h^2$ peaks in different frequency ranges as shown in Fig. \ref{Ogw_q}. The three models $p=1$, $p=2$ and $p=2/3$, with parameter sets listed in Table \ref{tb2}, generate similar gravitational wave spectrum which can be seen in various experiments. For example, the $\Omega_{\text{GW}} (k) h^2$ peak generated by parameter sets NG and SE can be seen in NANOGrav and SKA and may explain the stochastic process recently reported by NANOGrav collaboration. Similarly, the peak generated by parameter set TL can bee seen in future gravitational wave detectors such as, TaiJi, TianQin, LISA, AEDEG, DECIGO and BBO. Finally, the peak corresponding to parameter set ET lies in the sensitivity region of future experiments such as, AEDEG, DECIGO, BBO, CE and ET.

%Our model simultaneously explains the large scale CMB observations and leads to the generation of primordial black holes (PBHs) and scalar induced gravitational waves (SIGWs) at the small scales.

We can also compare the SIGW signal associated with the formation of primordial black holes (PBHs) to the recent NANOGrav results \cite{Arzoumanian:2020vkk}, which constrain the amplitude and slope of a stochastic process. The amplitude of the SIGW from \cite{KaiSmith20} is given as,
\begin{eqnarray}
\Omega_{\text{GW}}(f)=\Omega_{\text{gw}}^{\text{yr}}\left( \frac{f}{f_{\text{yr}}} \right)^{n_{\text{gw}}},
\end{eqnarray}
which allows direct comparison of our results to the NANOGrav bounds in the $\Omega_{\text{gw}}^{\text{yr}}-n_{\text{gw}}$ plane as shown by the dark (1-$\sigma$) and light (2-$\sigma$) shaded regions in Fig. \ref{YrGw}. We extract the amplitude and slope by comparing the amplitude at pivot scale $f_*=5.6\times 10^{-9}$ Hz and taking the logarithmic derivative of $\Omega_{\text{GW}}(f)$ at the pivot scale,
\begin{eqnarray}
n_{\text{gw}}&=& \left.\frac{d\log{\Omega_{\text{GW}}(f)}}{d\log{f}}\right|_{f=f_*},\\
\Omega_{\text{gw}}^{\text{yr}}&=&\Omega_{\text{GW}}(f_*)\left( \frac{f_{\text{yr}}}{f_*}\right)^{n_{\text{gw}}}.
\end{eqnarray}
Fig. \ref{YrGw} shows comparison of SIGW predictions, associated with the primordial black hole (PBH) formation, from hybrid inflation model with the constraints on the amplitude and tilt from NANOGrav \cite{Arzoumanian:2020vkk}. The curves are drawn for the models with $p = 1$, $p = 2$ and $p = 2/3$ using Gaussian peak function $K_g (\phi)$ (left panel) and step function $K_s (\phi)$ (right panel). It is evident that the predictions from all three models lie within the 1-$\sigma$ bound of NANOGrav. It is also important to emphasis that the broader peaks generated by Gaussian and step functions provide much better fitting to the NANOGrav bounds as compared to the sharp peaks of polynomial peak function.

\section{Summary} \label{sec_5}
To summarize, we have investigated production of primordial black holes (PBHs) and their associated induced secondary gravitational waves (SIGWs) using a background of hybrid inflation. The cosmological observables, such as the tensor-to-scalar ratio $r$ and the scalar spectral index $n_{s}$, are computed for various effective potentials. To produce the required abundance $Y_{pbh}$ for PBHs as a dark matter (DM) and their associated SIGWs, the curvature power spectrum is enhanced by seven order of magnitude, $P_\zeta \sim 0.01$ at the scale $k > \, \text{Mpc}^{-1}$ by employing a non-canonical kinetic energy term. %In particular, the field-dependent kinetic energy term can arise in many inflationary scenarios, such as, G-inflation, K-inflation, or general scalar-tensor theory of gravity. 
We have utilized a well known polynomial peak function $K_{q} (\phi)$ and employed two new functions; a Gaussian peak $K_{g} (\phi)$ and a step function $K_{s} (\phi)$, with a peak at $\phi_{p}$ in order to enhance the power spectrum at small scales. %The PBH mass and the frequency of secondary GWs are determined by the height $h$ and location of $\phi_{p}$, the peak value of power spectrum. 
The peak functions $K(\phi)$ induce an inflection point (with flat plateau) in the potential and effectively lead to ultra slow-roll inflation. However, the functions have a minor role away from the peak value where the usual slow-roll inflation is recovered, constrained by the CMB observations at large scales. Based on our analysis, it is quite possible that PBHs constitute most of the DM if the mass of PBH lies within the range  $(10^{-16} - 10^{-11}) \, M_{\odot}$ and $(1 - 11) \, M_{\odot}$. With the Gaussian and step function, wide mass range of PBH is realized $(10^{-16} - 11) \, M_{\odot}$ and the predictions of our model are in much better agreement with the experimental bounds. The frequencies of SIGWs, associated with the formation of PBHs, range from nHz to kHz, which can be detected by ground and space-based future GW observatories such as Square Kilometre Array (SKA), North American Nanohertz Observatory for Gravitational Waves (NANOGrav), Einstein Telescope (ET), Cosmic Explorer (CE), Laser Interferometer Gravitational-Wave Observatory (LIGO) O5, Laser Interferometer Space Antenna (LISA), Deci-hertz Interferometer Gravitational wave Observatory (DECIGO), Big Bang Observer (BBO), TaiJi, TianQin and Atomic Experiment for Dark Matter and Gravity Exploration in Space (AEDGE).  The evidence of stochastic process recently reported by analysis of 12.5 year NANOGrav data may be interpreted as SIGWs associated with the formation of PBHs. %The broad peaks in $\Omega_{GW}(f)$ generated by Gaussian and step functions in the nHz range fit well with the signal detected by the NANOGrav. 

\section*{Acknowledgments}
The authors would especially like to thank George K. Leontaris for very useful discussions, comments and revising the draft.

%%%%%%%%%%%%%%%%%%%%%%%%%%%%%%%%%

%%%%%%%%%%%%%%%%%%%%%%%%%%%%%%%


\begin{thebibliography}{99}
\bibliographystyle{unsrt}
%%%%%%%%%%%%%%%%%%%%%%%%%%%%%%%%%

%\cite{Bertone:2016nfn}
\bibitem{Bertone:2016nfn}
G.~Bertone and D.~Hooper,
%``History of dark matter,''
Rev. Mod. Phys. \textbf{90}, no.4, 045002 (2018)
doi:10.1103/RevModPhys.90.045002
[arXiv:1605.04909 [astro-ph.CO]].
%427 citations counted in INSPIRE as of 17 Jul 2021



\bibitem{Zeldovich:1966} 
  Y.~B.~Zel'dovich and I.~D.~Novikov, 
  The Hypothesis of Cores Retarded during Expansion and the Hot Cosmological Model,
  Sov.\ Astron. \ {\bf 10}, 602 (1967).

\bibitem{Hawking:1971ei} 
  S.~Hawking, 
  Gravitationally collapsed objects of very low mass,
  Mon.\ Not.\ Roy.\ Astron.\ Soc.\  {\bf 152}, 75 (1971).

\bibitem{Carr:1974nx} 
  B.~J.~Carr and S.~W.~Hawking, 
  Black holes in the early Universe,
  Mon.\ Not.\ Roy.\ Astron.\ Soc.\  {\bf 168}, 399 (1974).

\bibitem{Khlopov:2008qy} 
  M.~Y.~Khlopov, 
  Primordial Black Holes,
  Res.\ Astron.\ Astrophys.\  {\bf 10}, 495 (2010)
  %doi:10.1088/1674-4527/10/6/001
  [arXiv:0801.0116 [astro-ph]].

\bibitem{Sasaki:2018dmp} 
  M.~Sasaki, T.~Suyama, T.~Tanaka and S.~Yokoyama, 
  Primordial black holes -- perspectives in gravitational wave astronomy,
  Class.\ Quant.\ Grav.\  {\bf 35}, no. 6, 063001 (2018)
  %doi:10.1088/1361-6382/aaa7b4
  [arXiv:1801.05235 [astro-ph.CO]].

\bibitem{Carr:2016drx} 
  B.~Carr, F.~Kuhnel and M.~Sandstad, 
  Primordial Black Holes as Dark Matter,
  Phys.\ Rev.\ D {\bf 94}, no. 8, 083504 (2016)
  %doi:10.1103/PhysRevD.94.083504
  [arXiv:1607.06077 [astro-ph.CO]].

\bibitem{Carr:2018poi} 
  B.~Carr and F.~Kuhnel, 
  Primordial black holes with multimodal mass spectra,
  Phys.\ Rev.\ D {\bf 99}, no. 10, 103535 (2019)
  %doi:10.1103/PhysRevD.99.103535
  [arXiv:1811.06532 [astro-ph.CO]].

\bibitem{Hawking:1974rv} 
  S.~W.~Hawking, 
  Black hole explosions,
  Nature {\bf 248}, 30 (1974).
  %doi:10.1038/248030a0

\bibitem{MacGibbon:1991vc} 
  J.~H.~MacGibbon and B.~J.~Carr, 
  Cosmic rays from primordial black holes,
  Astrophys.\ J.\  {\bf 371}, 447 (1991).
  %doi:10.1086/169909

\bibitem{Carr:2009jm} 
  B.~J.~Carr, K.~Kohri, Y.~Sendouda and J.~Yokoyama, 
  New cosmological constraints on primordial black holes,
  Phys.\ Rev.\ D {\bf 81}, 104019 (2010)
  %doi:10.1103/PhysRevD.81.104019
  [arXiv:0912.5297 [astro-ph.CO]].

\bibitem{Niikura:2019kqi} 
  H.~Niikura, M.~Takada, S.~Yokoyama, T.~Sumi and S.~Masaki, 
  Constraints on Earth-mass primordial black holes from OGLE 5-year microlensing events,
  Phys.\ Rev.\ D {\bf 99}, no. 8, 083503 (2019)
  %doi:10.1103/PhysRevD.99.083503
  [arXiv:1901.07120 [astro-ph.CO]].

\bibitem{Carr:1997cn} 
  B.~J.~Carr and M.~Sakellariadou, 
  Dynamical constraints on dark compact objects,
  Astrophys.\ J.\  {\bf 516}, 195 (1999).
  %doi:10.1086/307071

\bibitem{Deng:2018wmy} 
  C.~M.~Deng, Y.~Cai, X.~F.~Wu and E.~W.~Liang, 
  Fast Radio Bursts From Primordial Black Hole Binaries Coalescence,
  Phys.\ Rev.\ D {\bf 98}, no. 12, 123016 (2018)
  %doi:10.1103/PhysRevD.98.123016
  [arXiv:1812.00113 [astro-ph.HE]].
  
  
  
  
\bibitem{GW1} B. P. Abbott \textit{et al.} (LIGO Scientific Collaboration and Virgo Collaboration), Phys. Rev. Lett. \textbf{116}, 061102 (2016).
\bibitem{GW2} B. P. Abbott \textit{et al.} (LIGO Scientific Collaboration and Virgo Collaboration), Phys. Rev. Lett. \textbf{116}, 241103 (2016).
\bibitem{GW3} B. P. Abbott \textit{et al.} (LIGO Scientific Collaboration and Virgo Collaboration), Phys. Rev. Lett. \textbf{118}, 221101 (2017).
\bibitem{GW4} B. P. Abbott \textit{et al.} (LIGO Scientific Collaboration and Virgo Collaboration), Astrophys. J. \textbf{851}, L35 (2017).
\bibitem{GW5} B. P. Abbott \textit{et al.} (LIGO Scientific Collaboration and Virgo Collaboration), Phys. Rev. Lett. \textbf{119}, 141101 (2017).


  
  
  
  
  
  

\bibitem{Sasaki:2016jop} 
  M.~Sasaki, T.~Suyama, T.~Tanaka and S.~Yokoyama, 
  Primordial Black Hole Scenario for the Gravitational-Wave Event GW150914,
  Phys.\ Rev.\ Lett.\  {\bf 117}, no. 6, 061101 (2016)
  Erratum: [Phys.\ Rev.\ Lett.\  {\bf 121}, no. 5, 059901 (2018)]
  %doi:10.1103/PhysRevLett.121.059901, 10.1103/PhysRevLett.117.061101
  [arXiv:1603.08338 [astro-ph.CO]].

\bibitem{Mandic:2016lcn} 
  V.~Mandic, S.~Bird and I.~Cholis, 
  Stochastic Gravitational-Wave Background due to Primordial Binary Black Hole Mergers,
  Phys.\ Rev.\ Lett.\  {\bf 117}, no. 20, 201102 (2016)
  %doi:10.1103/PhysRevLett.117.201102
  [arXiv:1608.06699 [astro-ph.CO]].
  
\bibitem{Wang:2016ana} 
S.~Wang, Y.~F.~Wang, Q.~G.~Huang and T.~G.~F.~Li, 
Constraints on the Primordial Black Hole Abundance from the First Advanced LIGO Observation Run Using the Stochastic Gravitational-Wave Background,
Phys.\ Rev.\ Lett.\  {\bf 120}, no. 19, 191102 (2018)
%doi:10.1103/PhysRevLett.120.191102
[arXiv:1610.08725 [astro-ph.CO]].


%\cite{Baumann:2007zm}
\bibitem{Baumann:2007zm}
D.~Baumann, P.~J.~Steinhardt, K.~Takahashi and K.~Ichiki,
%``Gravitational Wave Spectrum Induced by Primordial Scalar Perturbations,''
Phys. Rev. D \textbf{76}, 084019 (2007)
doi:10.1103/PhysRevD.76.084019
[arXiv:hep-th/0703290 [hep-th]].
%280 citations counted in INSPIRE as of 17 Jul 2021
\bibitem{Ananda:2006af}
K.~N.~Ananda, C.~Clarkson and D.~Wands,
%``The Cosmological gravitational wave background from primordial density perturbations,''
Phys. Rev. D \textbf{75}, 123518 (2007)
[arXiv:gr-qc/0612013 [gr-qc]].
%266 citations counted in INSPIRE as of 17 Jul 2021


\bibitem{Kohri:2018awv} 
  K.~Kohri and T.~Terada, 
  Semianalytic calculation of gravitational wave spectrum nonlinearly induced from primordial curvature perturbations,
  Phys.\ Rev.\ D {\bf 97}, no. 12, 123532 (2018)
  %doi:10.1103/PhysRevD.97.123532
  [arXiv:1804.08577 [gr-qc]].

\bibitem{Bartolo:2018rku} 
  N.~Bartolo, V.~De Luca, G.~Franciolini, M.~Peloso, D.~Racco and A.~Riotto, 
  Testing primordial black holes as dark matter with LISA,
  Phys.\ Rev.\ D {\bf 99}, no. 10, 103521 (2019)
  %doi:10.1103/PhysRevD.99.103521
  [arXiv:1810.12224 [astro-ph.CO]].

\bibitem{Cai:2019jah} 
  Y.~F.~Cai, C.~Chen, X.~Tong, D.~G.~Wang and S.~F.~Yan, 
  When Primordial Black Holes from Sound Speed Resonance Meet a Stochastic Background of Gravitational Waves,
  Phys.\ Rev.\ D {\bf 100}, no. 4, 043518 (2019)
  %doi:10.1103/PhysRevD.100.043518
  [arXiv:1902.08187 [astro-ph.CO]].
	
\bibitem{Cai:2018dig} 
R.~g.~Cai, S.~Pi and M.~Sasaki, 
Gravitational Waves Induced by non-Gaussian Scalar Perturbations,
Phys.\ Rev.\ Lett.\  {\bf 122}, no. 20, 201101 (2019)
%doi:10.1103/PhysRevLett.122.201101
[arXiv:1810.11000 [astro-ph.CO]].


%\cite{DeLuca:2020agl}
\bibitem{DeLuca:2020agl}
V.~De Luca, G.~Franciolini and A.~Riotto,
%``NANO-Grav Data Hints at Primordial Black Holes as Dark Matter,''
Phys. Rev. Lett. \textbf{126}, no.4, 041303 (2021)
[arXiv:2009.08268 [astro-ph.CO]].
%72 citations counted in INSPIRE as of 17 Jul 2021
K.~Inomata, M.~Kawasaki, K.~Mukaida and T.~T.~Yanagida,
%``NANO-Grav Results and LIGO-Virgo Primordial Black Holes in Axionlike Curvaton Models,''
Phys. Rev. Lett. \textbf{126}, no.13, 131301 (2021)
[arXiv:2011.01270 [astro-ph.CO]].
%17 citations counted in INSPIRE as of 17 Jul 2021
V.~Vaskonen and H.~Veerm\"ae,
%``Did NANO-Grav see a signal from primordial black hole formation?,''
Phys. Rev. Lett. \textbf{126}, no.5, 051303 (2021)
[arXiv:2009.07832 [astro-ph.CO]].
%66 citations counted in INSPIRE as of 17 Jul 2021





%\cite{Planck:2018jri}
\bibitem{Planck:2018jri}
Y.~Akrami \textit{et al.} [Planck],
%``Planck 2018 results. X. Constraints on inflation,''
Astron. Astrophys. \textbf{641}, A10 (2020)
[arXiv:1807.06211 [astro-ph.CO].
%1403 citations counted in INSPIRE as of 17 Jul 2021

%\cite{Lin:2020goi}
\bibitem{Lin:2020goi}
J.~Lin, Q.~Gao, Y.~Gong, Y.~Lu, C.~Zhang and F.~Zhang,
%``Primordial black holes and secondary gravitational waves from $k$ and $G$ inflation,''
Phys. Rev. D \textbf{101}, no.10, 103515 (2020)
[arXiv:2001.05909 [gr-qc]].
%36 citations counted in INSPIRE as of 17 Jul 2021






%\cite{Lu:2019sti}
\bibitem{Lu:2019sti}
Y.~Lu, Y.~Gong, Z.~Yi and F.~Zhang,
%``Constraints on primordial curvature perturbations from primordial black hole dark matter and secondary gravitational waves,''
JCAP \textbf{12}, 031 (2019)
[arXiv:1907.11896 [gr-qc]].
%30 citations counted in INSPIRE as of 17 Jul 2021


%\cite{Garcia-Bellido:2017aan}
\bibitem{Garcia-Bellido:2017aan}
J.~Garcia-Bellido, M.~Peloso and C.~Unal,
%``Gravitational Wave signatures of inflationary models from Primordial Black Hole Dark Matter,''
JCAP \textbf{09}, 013 (2017)
doi:10.1088/1475-7516/2017/09/013
[arXiv:1707.02441 [astro-ph.CO]].
%142 citations counted in INSPIRE as of 17 Jul 2021


%
\bibitem{Matarrese:1997ay}
S.~Matarrese, S.~Mollerach and M.~Bruni,
%``Second order perturbations of the Einstein-de Sitter universe,''
Phys. Rev. D \textbf{58}, 043504 (1998)
[arXiv:astro-ph/9707278 [astro-ph]].
%264 citations counted in INSPIRE as of 17 Jul 2021
%\cite{Ananda:2006af}



 %\cite{Smits:2008cf}
 \bibitem{Smits:2008cf}
 R.~Smits, M.~Kramer, B.~Stappers, D.~R.~Lorimer, J.~Cordes and A.~Faulkner,
 %``Pulsar searches and timing with the square kilometre array,''
 Astron. Astrophys. \textbf{493} (2009), 1161-1170
 %doi:10.1051/0004-6361:200810383
 [arXiv:0811.0211 [astro-ph]].
 %128 citations counted in INSPIRE as of 10 Feb 2022
 
   \bibitem{Arzoumanian:2020vkk}
   Zaven Arzoumanian(CRESST, Greenbelt and NASA, Goddard) et al. 
   NANOGrav Collaboration, 
   Astrophys. J. Lett. 905 (2020) 2, L34
   [arXiv: 2009.04496 [astro-ph.HE]].
 

   %\cite{Punturo:2010zz}
   \bibitem{Punturo:2010zz}
   M.~Punturo, M.~Abernathy, F.~Acernese, B.~Allen, N.~Andersson, K.~Arun, F.~Barone, B.~Barr, M.~Barsuglia and M.~Beker, \textit{et al.}
   %``The Einstein Telescope: A third-generation gravitational wave observatory,''
   Class. Quant. Grav. \textbf{27} (2010), 194002
   %doi:10.1088/0264-9381/27/19/194002
   %886 citations counted in INSPIRE as of 10 Feb 2022
   
    %\cite{LIGOScientific:2016wof}
    \bibitem{LIGOScientific:2016wof}
    B.~P.~Abbott \textit{et al.} [LIGO Scientific],
    %``Exploring the Sensitivity of Next Generation Gravitational Wave Detectors,''
    Class. Quant. Grav. \textbf{34}, no.4, 044001 (2017)
    %doi:10.1088/1361-6382/aa51f4
    [arXiv:1607.08697 [astro-ph.IM]].
    %632 citations counted in INSPIRE as of 11 Feb 2022
    %\cite{LIGOScientific:2019vic}
    
    \bibitem{LIGOScientific:2019vic}
    B.~P.~Abbott \textit{et al.} [LIGO Scientific and Virgo],
    %``Search for the isotropic stochastic background using data from Advanced LIGO\textquoteright{}s second observing run,''
    Phys. Rev. D \textbf{100}, no.6, 061101 (2019)
    %doi:10.1103/PhysRevD.100.061101
    [arXiv:1903.02886 [gr-qc]].
    %192 citations counted in INSPIRE as of 09 Feb 2022

%\cite{Ferdman:2010xq-Hobbs:2013aka}
\bibitem{Ferdman:2010xq}
R.~D.~Ferdman, R.~van Haasteren, C.~G.~Bassa, M.~Burgay, I.~Cognard, A.~Corongiu, N.~D'Amico, G.~Desvignes, J.~W.~T.~Hessels and G.~H.~Janssen, \textit{et al.}
%``The European Pulsar Timing Array: current efforts and a LEAP toward the future,''
Class. Quant. Grav. \textbf{27}, 084014 (2010)
doi:10.1088/0264-9381/27/8/084014
[arXiv:1003.3405 [astro-ph.HE]].
%76 citations counted in INSPIRE as of 17 Jul 2021
G.~Hobbs, A.~Archibald, Z.~Arzoumanian, D.~Backer, M.~Bailes, N.~D.~R.~Bhat, M.~Burgay, S.~Burke-Spolaor, D.~Champion and I.~Cognard, \textit{et al.}
%``The international pulsar timing array project: using pulsars as a gravitational wave detector,''
Class. Quant. Grav. \textbf{27}, 084013 (2010)
doi:10.1088/0264-9381/27/8/084013
[arXiv:0911.5206 [astro-ph.SR]].
%381 citations counted in INSPIRE as of 17 Jul 2021
M.~A.~McLaughlin,
%``The North American Nanohertz Observatory for Gravitational Waves,''
Class. Quant. Grav. \textbf{30}, 224008 (2013)
doi:10.1088/0264-9381/30/22/224008
[arXiv:1310.0758 [astro-ph.IM]].
%163 citations counted in INSPIRE as of 17 Jul 2021
%\cite{Hobbs:2013aka}
G.~Hobbs,
%``The Parkes Pulsar Timing Array,''
Class. Quant. Grav. \textbf{30}, 224007 (2013)
doi:10.1088/0264-9381/30/22/224007
[arXiv:1307.2629 [astro-ph.IM]].
%135 citations counted in INSPIRE as of 17 Jul 2021
K.~Danzmann,
%``LISA: An ESA cornerstone mission for a gravitational wave observatory,''
Class. Quant. Grav. \textbf{14}, 1399-1404 (1997)
doi:10.1088/0264-9381/14/6/002
%80 citations counted in INSPIRE as of 17 Jul 2021

%\cite{LISA:2017pwj}
\bibitem{LISA:2017pwj}
P.~Amaro-Seoane \textit{et al.} [LISA],
%``Laser Interferometer Space Antenna,''
[arXiv:1702.00786 [astro-ph.IM]].
%1204 citations counted in INSPIRE as of 17 Jul 2021

   %\cite{Seto:2001qf}
   \bibitem{Seto:2001qf}
   N.~Seto, S.~Kawamura and T.~Nakamura,
   %``Possibility of direct measurement of the acceleration of the universe using 0.1-Hz band laser interferometer gravitational wave antenna in space,''
   Phys. Rev. Lett. \textbf{87}, 221103 (2001)
   %doi:10.1103/PhysRevLett.87.221103
   [arXiv:astro-ph/0108011 [astro-ph]].
   %536 citations counted in INSPIRE as of 11 Feb 2022


      %\cite{Corbin:2005ny}
      \bibitem{Corbin:2005ny}
      V.~Corbin and N.~J.~Cornish,
      %``Detecting the cosmic gravitational wave background with the big bang observer,''
      Class. Quant. Grav. \textbf{23}, 2435-2446 (2006)
      %doi:10.1088/0264-9381/23/7/014
      [arXiv:gr-qc/0512039 [gr-qc]].
      %203 citations counted in INSPIRE as of 11 Feb 2

%\cite{Hu:2017mde}
\bibitem{Hu:2017mde}
W.~R.~Hu and Y.~L.~Wu,
%``The Taiji Program in Space for gravitational wave physics and the nature of gravity,''
Natl. Sci. Rev. \textbf{4}, no.5, 685-686 (2017)
doi:10.1093/nsr/nwx116
%126 citations counted in INSPIRE as of 17 Jul 2021

%\cite{TianQin:2015yph}
\bibitem{TianQin:2015yph}
J.~Luo \textit{et al.} [TianQin],
%``TianQin: a space-borne gravitational wave detector,''
Class. Quant. Grav. \textbf{33}, no.3, 035010 (2016)
doi:10.1088/0264-9381/33/3/035010
[arXiv:1512.02076 [astro-ph.IM]].
%372 citations counted in INSPIRE as of 17 Jul 2021

  %\cite{AEDGE:2019nxb}
  \bibitem{AEDGE:2019nxb}
  Y.~A.~El-Neaj \textit{et al.} [AEDGE],
  %``AEDGE: Atomic Experiment for Dark Matter and Gravity Exploration in Space,''
  EPJ Quant. Technol. \textbf{7}, 6 (2020)
  %doi:10.1140/epjqt/s40507-020-0080-0
  [arXiv:1908.00802 [gr-qc]].
  %101 citations counted in INSPIRE as of 11 Feb 2022

%\cite{Ahmed:2014cma}
\bibitem{Ahmed:2014cma}
W.~Ahmed, O.~Ishaque and M.~U.~Rehman,
%``Quantum Smearing in Hybrid Inflation with Chaotic Potentials,''
Int. J. Mod. Phys. D \textbf{25}, no.03, 1650035 (2016)
doi:10.1142/S0218271816500358
[arXiv:1501.00173 [hep-ph]].
%2 citations counted in INSPIRE as of 17 Jul 2021


%\cite{Coleman:1973jx}
\bibitem{Coleman:1973jx} 
  S.~R.~Coleman and E.~J.~Weinberg,
  ``Radiative Corrections as the Origin of Spontaneous Symmetry Breaking,''
  Phys.\ Rev.\ D {\bf 7}, 1888 (1973).
  %%CITATION = PHRVA,D7,1888;%%
  %3321 citations counted in INSPIRE as of 28 Dec 2014 

%\cite{NeferSenoguz:2008nn}
\bibitem{NeferSenoguz:2008nn} 
  V.~N.~Senoguz and Q.~Shafi,
  ``Chaotic inflation, radiative corrections and precision cosmology,''
  Phys.\ Lett.\ B {\bf 668}, 6 (2008)
  [arXiv:0806.2798 [hep-ph]].
  %%CITATION = ARXIV:0806.2798;%%
  %28 citations counted in INSPIRE as of 28 Dec 2014
  
  
  
  %\cite{Dvali:1994ms}
  \bibitem{Dvali:1994ms}
  G.~R.~Dvali, Q.~Shafi and R.~K.~Schaefer,
  ``Large scale structure and supersymmetric inflation without fine tuning,''
  Phys.\ Rev.\ Lett.\  {\bf 73}, 1886 (1994)
  [arXiv:hep-ph/9406319].
  %%CITATION = PRLTA,73,1886;%%

%\cite{Kobayashi:2010cm}
\bibitem{Kobayashi:2010cm}
T.~Kobayashi, M.~Yamaguchi and J.~Yokoyama,
%``G-inflation: Inflation driven by the Galileon field,''
Phys. Rev. Lett. \textbf{105}, 231302 (2010)
[arXiv:1008.0603 [hep-th]].
%408 citations counted in INSPIRE as of 17 Jul 2021

%\cite{Garriga:1999vw}
\bibitem{Garriga:1999vw}
J.~Garriga and V.~F.~Mukhanov,
%``Perturbations in k-inflation,''
Phys. Lett. B \textbf{458}, 219-225 (1999)
[arXiv:hep-th/9904176 [hep-th]].
%1012 citations counted in INSPIRE as of 17 Jul 2021

\bibitem{TasiCosmology2018}
D.~Baumann, TASI Lectures on Primordial Cosmology,
[arXiv:1807.03098 [hep-th].

\bibitem{Jennifer}
Jennifer A. Adams, Graham G. Ross and Subir Sarkar,
Phys. Lett. B 391 (1997) 271-280,
[arXiv: 9608336 [hep-ph]].
Jennifer Adams, Bevan Cresswell and Richard Easther,
Phys.Rev.D 64 (2001) 123514,
[arXiv: 0102236 [astro-ph]].


 
  %\cite{Carr:1975qj}
\bibitem{Carr:1975qj}
B.~J.~Carr,
%``The Primordial black hole mass spectrum,''
Astrophys. J. \textbf{201}, 1-19 (1975)
doi:10.1086/153853
%791 citations counted in INSPIRE as of 30 Jul 2021
  
  
  
  
  %\cite{Ozsoy:2018flq}
\bibitem{Ozsoy:2018flq}
O.~\"Ozsoy, S.~Parameswaran, G.~Tasinato and I.~Zavala,
%``Mechanisms for Primordial Black Hole Production in String Theory,''
JCAP \textbf{07}, 005 (2018)
doi:10.1088/1475-7516/2018/07/005
[arXiv:1803.07626 [hep-th]].
%77 citations counted in INSPIRE as of 30 Jul 2021


%\cite{Tada:2019amh}
\bibitem{Tada:2019amh}
Y.~Tada and S.~Yokoyama,
%``Primordial black hole tower: Dark matter, earth-mass, and LIGO black holes,''
Phys. Rev. D \textbf{100}, no.2, 023537 (2019)
doi:10.1103/PhysRevD.100.023537
[arXiv:1904.10298 [astro-ph.CO]].
%37 citations counted in INSPIRE as of 30 Jul 2021
  

%\cite{Planck:2018vyg}
\bibitem{Planck:2018vyg}
N.~Aghanim \textit{et al.} [Planck],
%``Planck 2018 results. VI. Cosmological parameters,''
Astron. Astrophys. \textbf{641}, A6 (2020)
doi:10.1051/0004-6361/201833910
[arXiv:1807.06209 [astro-ph.CO]].
%5551 citations counted in INSPIRE as of 30 Jul 2021


%\cite{Musco:2012au}
\bibitem{Musco:2012au}
I.~Musco and J.~C.~Miller,
%``Primordial black hole formation in the early universe: critical behaviour and self-similarity,''
Class. Quant. Grav. \textbf{30}, 145009 (2013)
doi:10.1088/0264-9381/30/14/145009
[arXiv:1201.2379 [gr-qc]].
%145 citations counted in INSPIRE as of 30 Jul 2021
\bibitem{bardeen80}
Bardeen, J. M. (1980) Gauge-invariant cosmological perturbations, Phys. Rev. D 22, 1882-1905.

\bibitem{Terada}
K. Inomata and T. Terada, Phys. Rev. D101, 023523
(2020), arXiv:1912.00785 [gr-qc].



\bibitem{KaiSmith20}
Wilfried Buchmuller, Valerie Domcke and Kai Schmitz,
Phys.Lett.B 811 (2020) 135914, 
[arXiv:1201.2379 [gr-qc]].






%%%%%%%%%%%%%%%%%%%%%%%%%%%%%%%
\end{thebibliography}
\end{document}